\pdfoutput=1
\documentclass{ws-rv9x6}
\usepackage{subfigure}   
\usepackage{ws-rv-thm}   
\usepackage[square]{ws-rv-van}   
\makeindex

\usepackage{hyperref}
\hypersetup{
    colorlinks,%
    citecolor=blue,%
    filecolor=black,%
    linkcolor=blue,%
    urlcolor=blue,%
    pdftitle={Artificial Intelligence for High Energy Physics - Boosted decision trees},%
    pdfauthor={Yann Coadou}
}

\usepackage[utf8]{inputenc}
\usepackage{xspace}
\newcommand{\dt}{decision tree\xspace}
\newcommand{\dts}{decision trees\xspace}

\newcommand{\DTs}{Decision trees\xspace}
\newcommand{\bdt}{boosted decision tree\xspace}
\newcommand{\bdts}{boosted decision trees\xspace}
\newcommand{\BDT}{Boosted decision tree\xspace}
\newcommand{\BDTs}{Boosted decision trees\xspace}

\begin{document}

\setcounter{secnumdepth}{3}
\setcounter{tocdepth}{3}

\chapter*{\BDTs}\label{ra_ch1}

\author[Y. Coadou]{Yann Coadou}

\address{Centre de physique des particules de Marseille (CPPM),\\ Aix Marseille Université, CNRS/IN2P3, Marseille, France \\
coadou@cppm.in2p3.fr}

\begin{abstract}
  \BDTs are a very powerful machine learning technique. After
  introducing specific concepts of machine learning in the high-energy
  physics context and describing ways to quantify the performance and
  training quality of classifiers, \dts are described. Some of their
  shortcomings are then mitigated with ensemble learning, using
  boosting algorithms, in particular AdaBoost and gradient
  boosting. Examples from high-energy physics and software used are
  also presented.
  \vfill
  \textit{To appear in \href{https://doi.org/10.1142/12200}{Artificial Intelligence for High Energy Physics}, P.~Calafiura, D.~Rousseau and K.~Terao, eds. (World Scientific Publishing, 2022)}
\end{abstract}

\body

\clearpage

\tableofcontents

\newpage
\section{Introduction}
\label{sec:Introduction}
\DTs are a machine learning technique that appeared in the mid-1980's
and are still the subject of advanced studies in the field. Because it
is a sophisticated supervised multivariate technique, learning from
examples, it is important to remember that before applying it to real
data (e.g. collisions from a high-energy physics experiment), it is crucial to have a good understanding of the data and of
the physics model used to describe them (simulated samples, reconstruction and identification efficiencies, etc.). Any discrepancy between the real data and
physics model (that is, features in the data that are not reproduced by the physics model because the simulation is incorrect or because the real data were not properly groomed) will provide an artificial separation that the \dts\ will use,
misleading the analyser. The hard (and interesting) part of the analysis
is in building the proper physics model, not in `just' extracting the
signal. But once this is properly done, \dts\ (and especially their
boosted versions) provide a very powerful tool to increase the
significance of any analysis.

Ever since their first use by the MiniBooNe collaboration for analysis
and particle identification~\cite{Miniboone1,Miniboone2} and by the D0
experiment for the first evidence of single top quark
production~\cite{D01,D02}, \bdts have been a primary tool in high
energy physics to increase the discovery potential and measurement
precision of experiments, in particular at the Tevatron and at the
LHC. They are still highly relevant (and highly performing) in 2021,
even though deep neural networks are becoming a serious contender.

As this is the first chapter of this book, some of the basic concepts
useful in the context of high-energy physics when using most
techniques presented in this and other chapters are summarised in
\sref{sec:Performance}.
\Sref{sec:DT} explains how a \dt\ is constructed, what parameters
can influence its development and what its intrinsic limitations are.
One possible extension of \dts, boosting, is described in detail
in \sref{sec:BDT}, and other techniques trying to reach the same goal
as boosting are presented in \sref{sec:Others}. Popular software
implementations are introduced in \sref{sec:Soft}, before reaching
conclusions in \sref{sec:Conclusion}.

\section{Specificity of high-energy physics}
\label{sec:Performance}

All techniques presented in this book need to learn from
examples. After a short list of definitions to have a common language
between the physicist and the computer scientist in
\sref{sec:terminology}, several training strategies are presented in
\sref{sec:samplesplit}, as well as how to deal with the samples to
minimise training bias and maximise statistical power. In order to
properly assess the performance of a classifier, cross-validation is
introduced in \sref{sec:crossval}. \Sref{sec:usingML} describes
typical usage of machine learning algorithms in high-energy
physics. Several figures of merit are described in \sref{sec:fom} and
overtraining is addressed in \sref{sec:overtraining}.

\subsection{Terminology}
\label{sec:terminology}
Here are a few terms that take on different meanings in a high-energy physics or machine learning context.
\begin{description}
\item[Event] All information collected during a collision inside a detector, or reproduced from a Monte Carlo simulation of such collisions (equivalent to a `sample' in machine learning literature).
\item[Sample] A collection of events, a dataset.
\item[Variable] A property of the event or of one of its constituents (`feature' in machine learning)
\item[Cut] To cut on a variable is to apply a threshold on this variable and keep only events satisfying this condition. A cut-based analysis is applying such thresholds on several variables to select events.
\item[Event weight] In high-energy physics events usually have an
  associated weight, which depends on how many events were generated
  (relating to the process cross section and collected luminosity) and
  various corrections applied to simulations to account for
  differences between data and Monte Carlo predictions (jet energy
  scale or object identification efficiency are such weights). When
  using machine learning techniques all events are often treated equal
  by default. It is therefore important for the physicist to make sure
  to give the proper initial weight to all its input events. Then
  machine learning algorithms may internally reweight the events for
  their own purpose, but the starting point will correspond to the
  physical distributions. The concept is similar to importance
  weighting in machine learning, where events are given a larger
  weight to account, for instance, for their scarcity in the training
  sample.
\end{description}

\subsection{Splitting samples for training}
\label{sec:samplesplit}
\DTs, as many of the techniques presented in this book, belong to the
class of algorithms using supervised learning: during training, the
classifier is presented only with events for which it knows features
(discriminating variables) and class label (for instance in the binary
case, whether the event is signal or background).

In order to not introduce bias, it is
important to use an independent set of events during training, events
that are then not used when performing a measurement. The usual
approach is to split the dataset in three parts: a training sample
from which to learn the task, a validation sample to evaluate
performance and possibly optimise the classifier hyperparameters, and
a testing sample for the actual measurement. In high-energy physics,
simulated Monte Carlo events are often used for these three samples,
and the performance on the testing sample is compared to that on data
collected from the detector (never seen during
training). Discrepancies between testing sample and data introduce a
potential pitfall, that can be addressed with transfer learning and domain adaptation~\cite{BenDavid2009}.

In general labelled data are `expensive' to produce: hiring people to
label images or translate speech, collecting X-ray images and medical
diagnosis, etc. In high-energy physics very accurate, though not
perfect, event generators and detector simulators are
available. Models can be trained on the samples they provide, which are
however quite costly in resources so that they should be
used with parsimony. At the same time an increasing training set size is often
associated with improved classifier performance. Monte
Carlo samples can be split in half, one half for
training (holding out part of this dataset for validation) and one for
testing. By doing this half of the sample is `wasted', not used for either
training or testing, decreasing the quality of the training and of the
measurement. The use of the sample can be maximised by
performing two trainings: train the same classifier on the two halves
(say, one on events with an even event number and one on events with
an odd event number), and when testing, apply the classifier which did
not see the event during training (so, the one trained on odd events is
applied on even ones, and vice versa). The concept can be generalised
to any number of splits, increasing the number of trained classifiers,
each of them using a larger fraction of the available dataset for
training.

\subsection{Cross-validation}
\label{sec:crossval}
Training machine learning algorithms is usually a stochastic problem,
the randomness coming from the training sample content, the
optimisation process or the technique itself. This means that when
training only once, there is a possibility to obtain an `abnormal'
result by chance, too good or too bad compared to what could be
expected. In high-energy physics some training samples may
be limited in size and become very sensitive to this issue.
To get a proper estimate of the mean performance
and associated uncertainty (the variability of the algorithm output, originating from the training procedure), it may be better to perform several
trainings. This principle was introduced with the so-called $K$-fold
cross-validation, originally for \dts~\cite{Breiman}. After dividing a training
sample $\mathcal{L}$ into $K$ subsets of equal size,
$\mathcal{L}=\bigcup_{k=1..K}\mathcal{L}_k$, a
classifier $T_k$ is trained on the $\mathcal{L}-\mathcal{L}_k$ sample and tested
on $\mathcal{L}_k$. This produces $K$ classifiers, from which
the mean performance and associated uncertainty is extracted. It helps in
choosing the best model (each being tested with cross-validation),
rather than relying on a single training for each model (which may or
may not have an upward or downward performance fluctuation). Once the
model is chosen, it can be retrained on the full (larger) training
set, assuming its performance should approach the observed mean
performance.

\subsection{Using machine learning}
\label{sec:usingML}
\begin{figure}
  \centerline{
    \subfigure[]
    {\includegraphics[height=4.9cm]{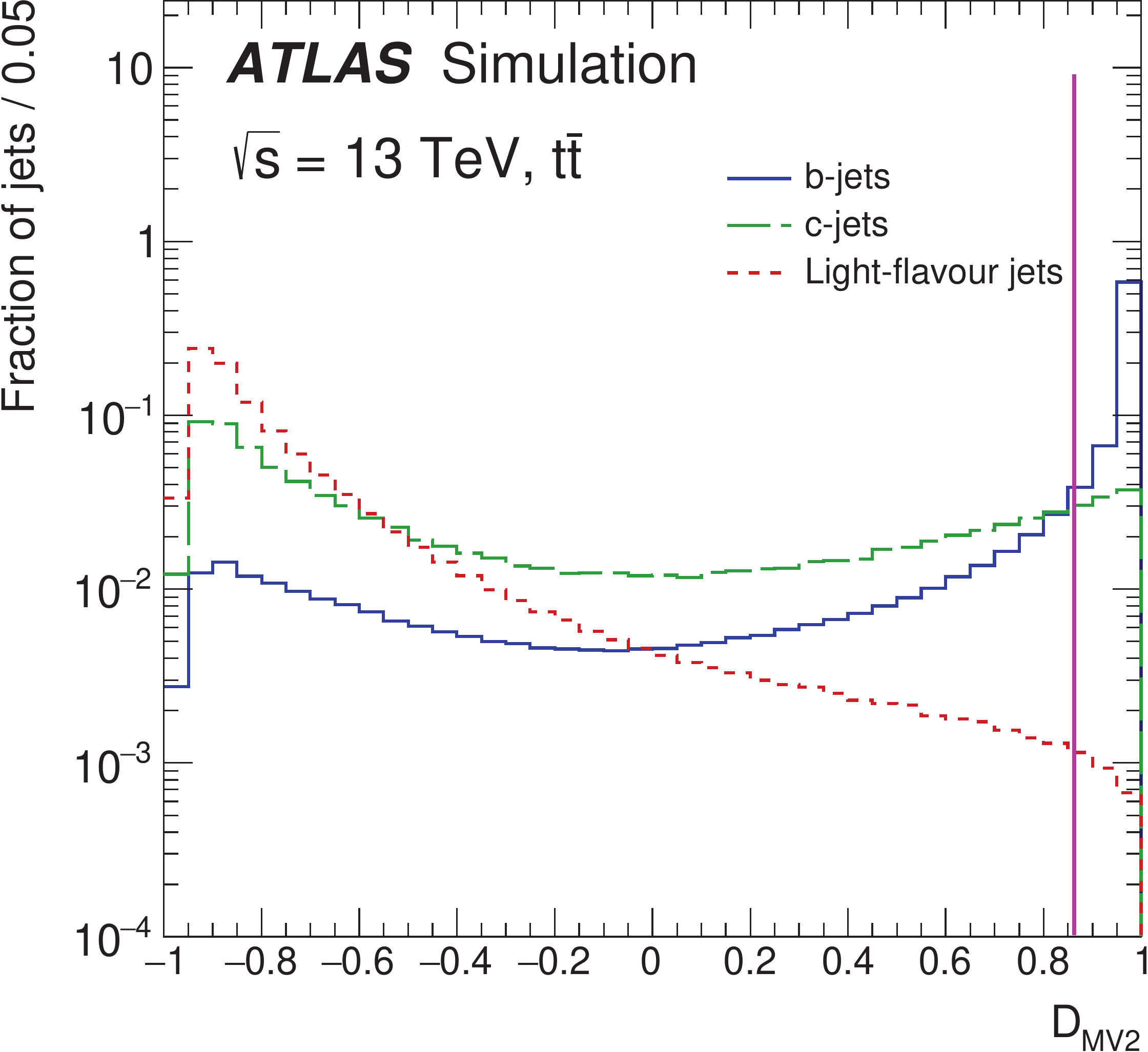}\label{fig:BDTusage:mv2}}
    \hspace*{20pt}
    \subfigure[]
    {\includegraphics[height=4.9cm]{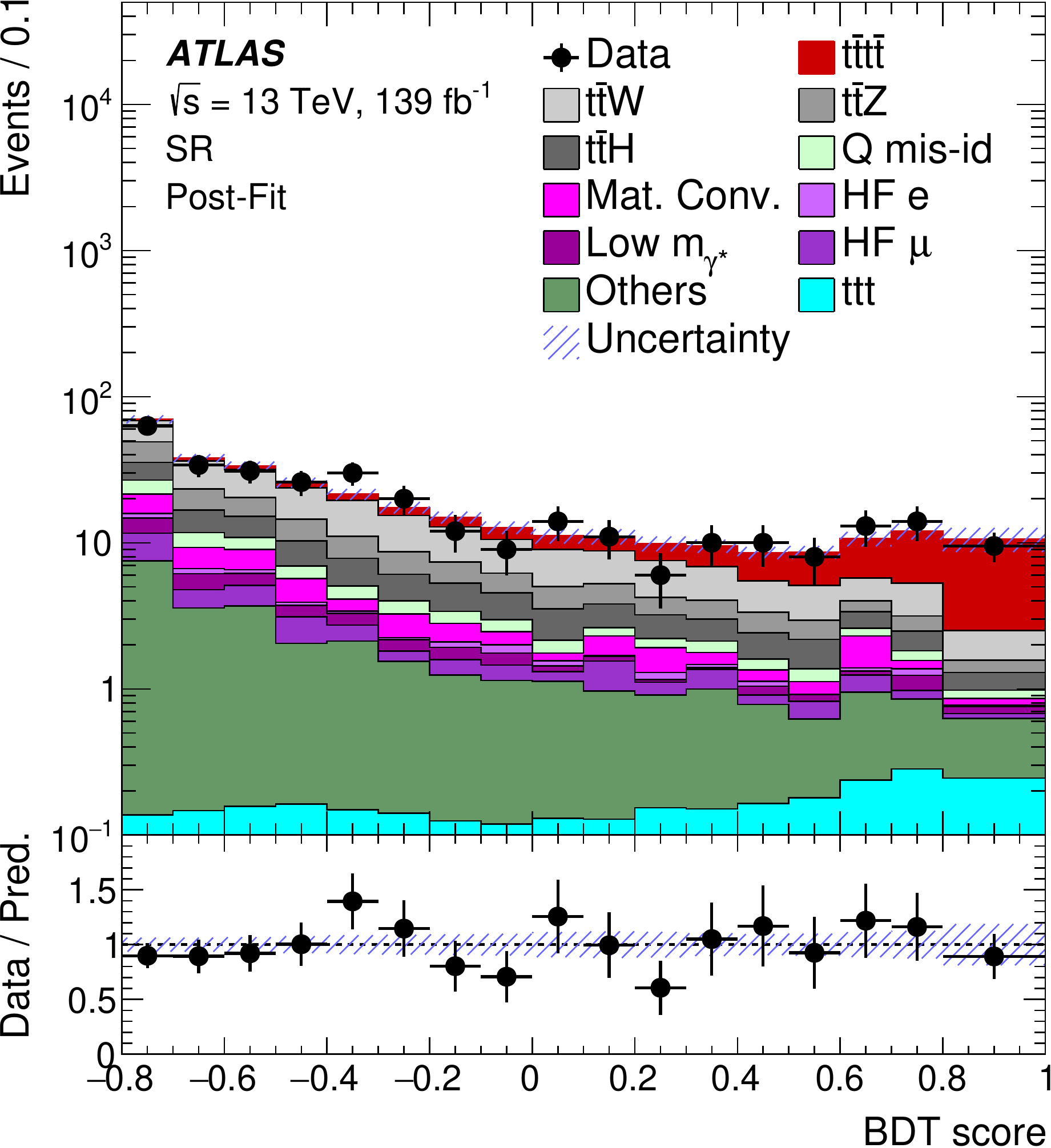}\label{fig:BDTusage:tttt}}
  }
  \caption{(a) Output of the \bdt used to identify jets originating
    from $b$-quarks in ATLAS~\cite{FTAG-2018-01}. (b) \BDT output used
    in a fit between data and physics model to extract the
    $t\bar tt\bar t$ signal~\cite{TOPQ-2018-05}.}
  \label{fig:BDTusage}
\end{figure}

This book describes various ways of using machine learning in high
energy physics to accomplish many different tasks. \BDTs are mostly
used to separate a rare signal from a large background in physics
analyses or to identify physics objects in the detector (see several
use cases in \sref{sec:UseCases}). In practice these results are
obtained in two ways. By applying a threshold on the \bdt output a
region or working point can be defined, as shown in
\fref{fig:BDTusage:mv2}: cutting at 0.83 defines a $b$-tagging working
point with 70\% efficiency on $b$-jets and a rejection factor (defined as the inverse of efficiency) of 313
(8) against light-flavoured jets ($c$-jets)~\cite{FTAG-2018-01}. The
second approach consists in using the shape of the \bdt output as the
discriminating variable for the final analysis. As an example, in
\fref{fig:BDTusage:tttt} the bins of `BDT score' for all components of
the physics model are included in a binned likelihood fit to the
data. The low score values help constrain the background, and the high
score bins reveal the need of the signal contribution to match the
data, leading to the first evidence for the production of
$t\bar tt\bar t$ in ATLAS~\cite{TOPQ-2018-05}.

\subsection{Figures of merit}
\label{sec:fom}
It is nowadays very easy, in just a few lines, to write the
code to train and apply various machine learning
algorithms, with several software options on the market (see \sref{sec:Soft}). The lengthy
part is more in the design and optimisation of the model itself (that is, what algorithm, structure, hyperparameters to put in these few lines), and
how to pick the best one. Several measures that are commonly used, in particular in high-energy physics, are
presented below.

\subsubsection{ROC curve and area under the curve}
\label{sec:ROC}
\begin{figure}
  \centerline{\includegraphics[width=.5\textwidth]{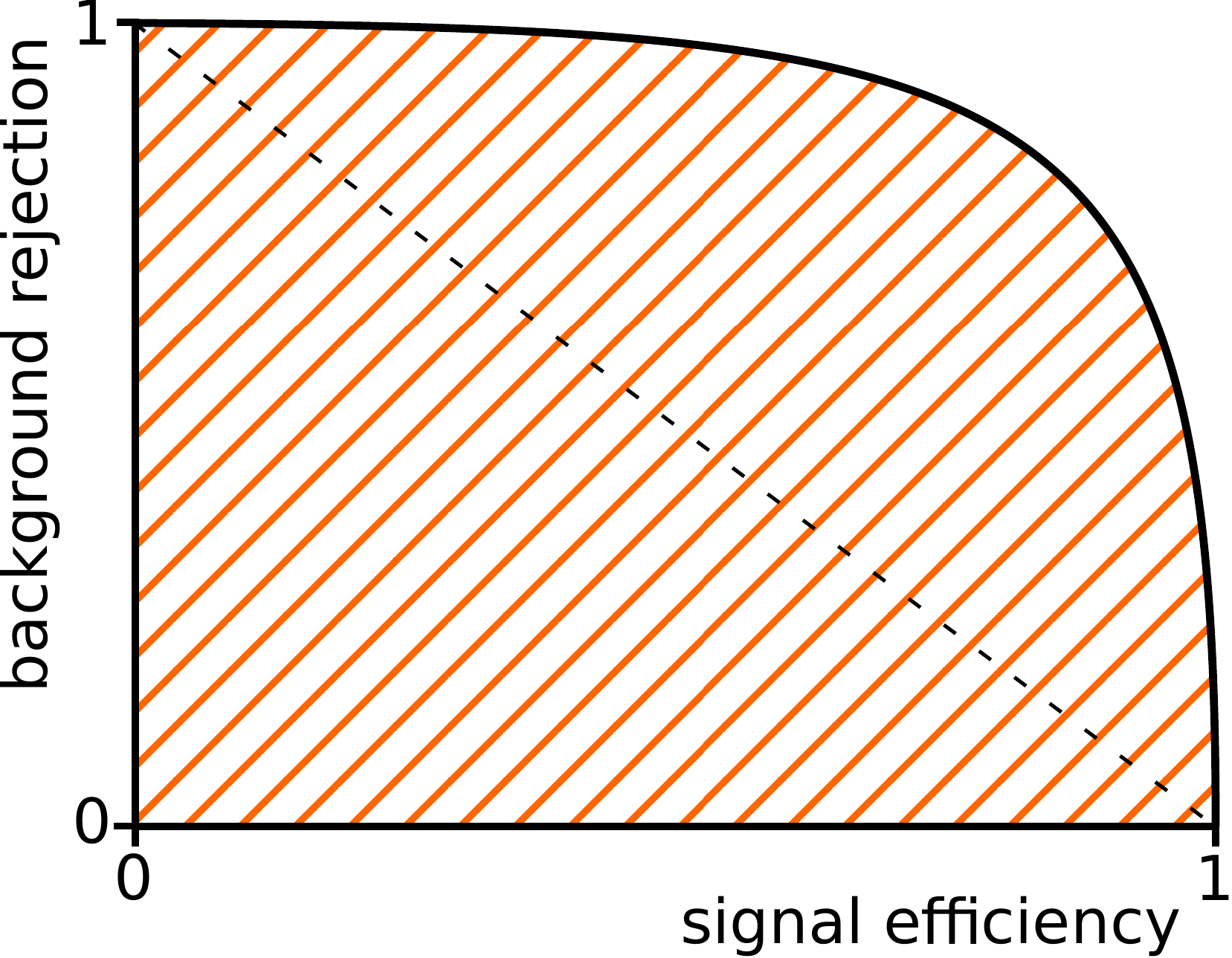}}
  \caption{Example ROC curve. The hatched area is the area under the
    curve. The dashed line corresponds to random guessing.}
  \label{fig:fom:roc}
\end{figure}

The receiver operating characteristic curve, or ROC curve, is a
representation of the capacity of a binary classifier to separate the
two classes, as its discrimination threshold is varied. It is plotting
the true positive rate (or recall, a measure of the proportion of
actual positives that are correctly identified as such) against the
false positive rate (or fall-out, actual negatives improperly
identified as positive), obtained when scanning the classifier
output. In the context of signal and background, it shows signal
efficiency versus background efficiency (or background rejection,
defined as $1-\text{efficiency}$). An example is shown in
\fref{fig:fom:roc}. In this convention, the better the classifier, the
closer the curve is to the top right corner. The dashed line in the
middle represents the performance of a classifier that is randomly
guessing, rejecting or accepting 50\% of signal and background in all
cases. This is the worst achievable performance.

To compare ROC curves between classifiers, the area
under the curve, AUC (hatched area in \fref{fig:fom:roc}) can be computed. Perfect
separation gives an AUC of one, while random guessing corresponds to
an AUC of 0.5.

A single number summary is of course practical, but hides details of
the ROC curves being compared. If one ROC curve is systematically above the other, its
AUC is larger and reflects better performance across the board. But if
two ROC curves cross each other, then the interpretation of the AUC is
more tricky: depending on the usage of the classifier, a higher curve
at high background rejection may be more interesting than one at high
efficiency for instance, so how to interpret the AUC is up to the
analyser. To partially account for this effect it is also possible to
compute the AUC only above a certain threshold.

\subsubsection{Significance}
\label{sec:significance}
In a physics analysis, the AUC is rarely the number of interest to
optimise. It is more typical to aim for the best cross-section
significance $\frac{s}{\sqrt{s+b}}$ or excess significance
$\frac{s}{\sqrt{b}}$, where $s$ ($b$) is the sum of weights (see \sref{sec:terminology}) of signal
(background) events. With $n$ events in data, the observed
significance is obtained by replacing $s$ by $n-b$. Given a machine
learning algorithm output, typically in the range $[0,1]$ or $[-1,1]$,
as is done when producing the ROC curve, $s$ and $b$ are computed 
above a threshold on the discriminant output, scanning its full range.
It usually goes
through a maximum towards high output values, before decreasing when
statistics become too small. This maximum significance corresponds to
the optimal value on which to cut on the discriminant to get the best
possible analysis.

This simple-minded formula is very popular in high-energy physics but
has shortcomings, and a refined version (counting experiment supposing a single Poisson distributed value, with known background) gives the approximate median
significance~\cite{Cowan:2010js}:
\[ \text{AMS} = \sqrt{2\left((s+b)\ln\left(1+\frac{s}{b}\right) - s\right)}. \]

Expanding the logarithm in $s/b$ leads back to the previous formula,
qualifying the validity of the approximation (requires $s\ll b$):
\[ \text{AMS} = \frac{s}{\sqrt{b}}\left(1+\mathcal{O}(s/b)\right). \]

Optimising the AMS corresponds to optimising the ROC curve, focusing
on the region with very high background rejection. This is the typical
regime of a physics analysis.

There is usually an uncertainty on the background, which affects the
significance. To extract their final results, modern analyses rely on
advanced statistical models with a complex machinery (usually based on
the RooStat framework~\cite{RooStat}) accounting for all
possible systematic effects. Running this whole infrastructure during
machine learning training optimisation is usually prohibitive (complexity, CPU cost),
so a simpler proxy to the analysis performance measure is necessary.

The simplest way to account partially for background uncertainty
($\sigma_b \equiv ||b-b_\text{syst}||$) is to replace $\sqrt{b}$ by
the quadratic sum of $\sqrt{b}$ and $\sigma_b$:
\[ \frac{s}{\sqrt{b+\sigma_b^2}}. \]

A refined version of the AMS can also take into account the background
uncertainty~\cite{pmlr-v42-cowa14}:
\[  \text{AMS}_1 = \sqrt{2\left((s+b)\ln\frac{s+b}{b_0} - s - b + b_0 \right) + \frac{(b-b_0)^2}{\sigma^2_b}},
\]
\[ \text{with}\quad b_0 = \frac{1}{2}\left( b - \sigma^2_b + \sqrt{(b - \sigma^2_b)^2 + 4(s + b)\sigma^2_b} \right). \]

Expanding in powers of $s/b$ and $\sigma^2_b/b$ gives back the simpler
formula:

\[ \frac{s}{\sqrt{b+\sigma_b^2}}\left(1+\mathcal{O}(s/b)+\mathcal{O}(\sigma_b^2/b)\right). \]

Finally, to account for the shape of the discriminant rather than only
choosing the best cut in a counting experiment, it is possible to
replace the global counts $s$, $b$ and $\sigma_b$ by their counts in
each bin, summing up contributions of $N$ bins of discriminant output:
\[  \text{AMS}_1^\text{sum} = \sqrt{\sum_i^N \left(2\left((s_i+b_i)\ln\frac{s_i+b_i}{b_{0i}} - s_i -b_i + b_{0i} \right) + \frac{(b_i-b_{0i})^2}{\sigma^2_{bi}}\right)}, \]

\[  b_{0i} = \frac{1}{2}\left(b_i - \sigma^2_{bi} + \sqrt{(b_i - \sigma^2_{bi})^2 + 4(s_i + b_i)\sigma^2_{bi}} \right). \]

\subsection{Controlling overtraining}
\label{sec:overtraining}
Overtraining is what happens when a classifier learns too much about
the specific details of the training sample, while these features are
not representative of the underlying distributions. It may then be
targeting noise, or misrepresent regions with too little statistics to
train on. When applying such a classifier on the testing sample, its
performance will be worse than that of a classifier immune to
this issue, because it does not generalise well. It should be noted that what is often called overtraining here and in the following, in accordance with high-energy physics usage, is usually referred to as overfitting in the machine learning community.
This is the so-called
bias--variance trade-off~\cite{hastie2001elements}: it is difficult to
minimise both the bias (the difference between the prediction of the
model and the correct value it tries to predict) and variance (the
variability of the model prediction for a given event, when considering multiple realisations of this same model). Increasing
model complexity lowers the bias while increasing variance.

A particular type of overtraining is very easy to avoid, by following
good practices from \sref{sec:samplesplit}: never use
training events when making the final measurement, which has to be
performed on an independent set of events, never seen during
training. Otherwise the performance will be artificially enhanced on
the `testing' sample and comparisons with the application to data will
be impossible (or worse if not noticed).

Several techniques exist to mitigate overtraining, generically referred to as
regularisation. They typically add a penalty for complexity to the
loss function that is minimised during training (the function that maps each event to a real number quantifying the difference between the predicted and true classes or values). With classifier $f$ and loss function $L$, a regularisation term $R(f)$ is added to the loss function, which becomes $L(f) + \lambda R(f)$, 
where $\lambda$ is a parameter controlling the importance of the regularisation term.
This will favour
simpler models (increasingly simpler with larger values of $\lambda$, with the risk of underfitting with too much regularisation), less susceptible to overtraining. $R(f)$ can take various
forms, like L1 (L2) regularisation based on the sum of weights (sum of
squared weights) used to describe neural networks, or the number and depth of trees (see
\sref{sec:prune}). Sparsity (setting many weights to zero~\cite{sparsity}) and dropout
(randomly dropping out nodes during training~\cite{dropout}) are more recent very
effective approaches for neural networks. Ensemble learning (see
\sref{sec:average} and \sref{sec:Others}) is another approach.

\begin{figure}
  \centerline{
    \subfigure[]
    {\includegraphics[width=.45\textwidth]{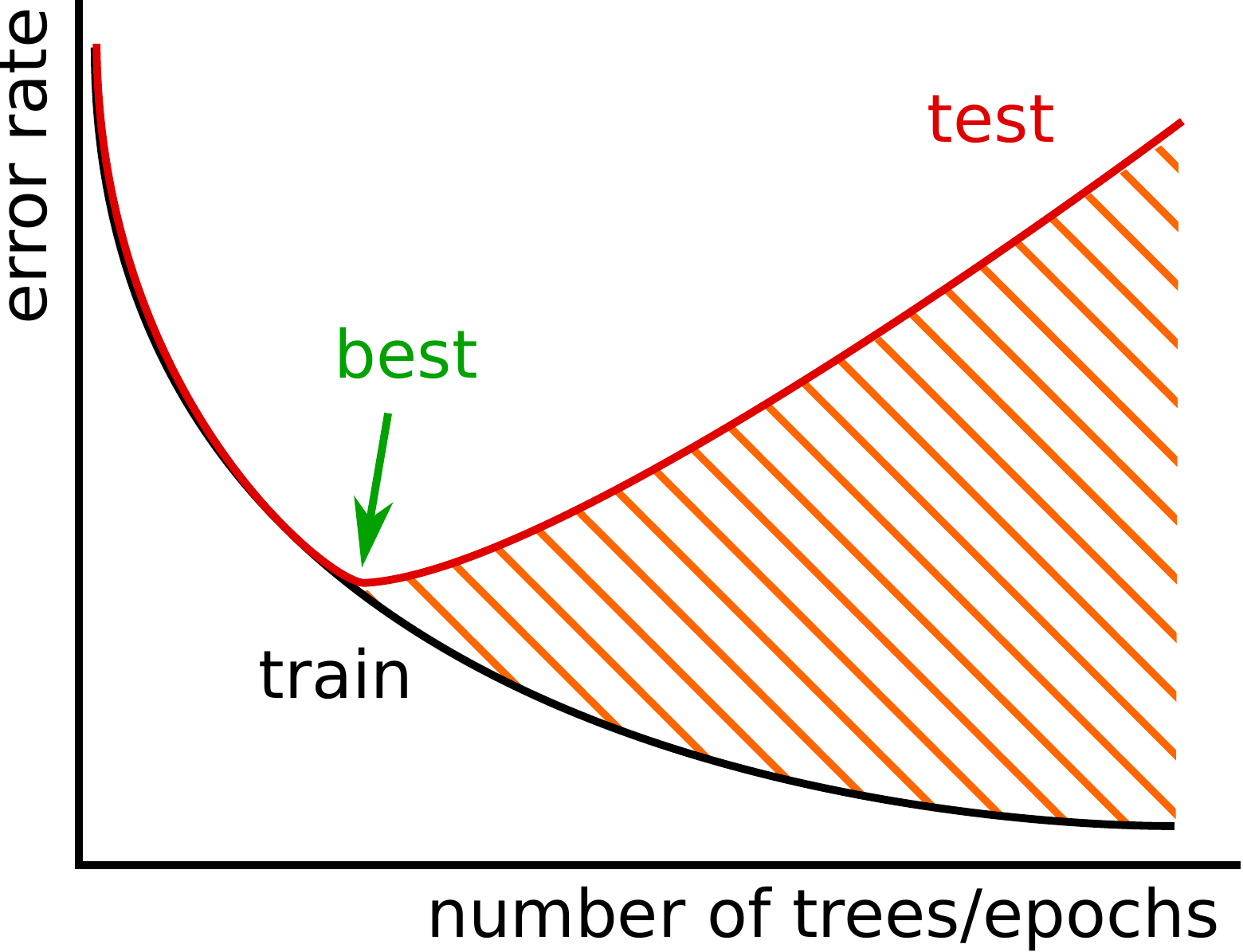}\label{fig:overtraining:usual}}
    \hspace*{10pt}
    \subfigure[]
    {\includegraphics[width=.45\textwidth]{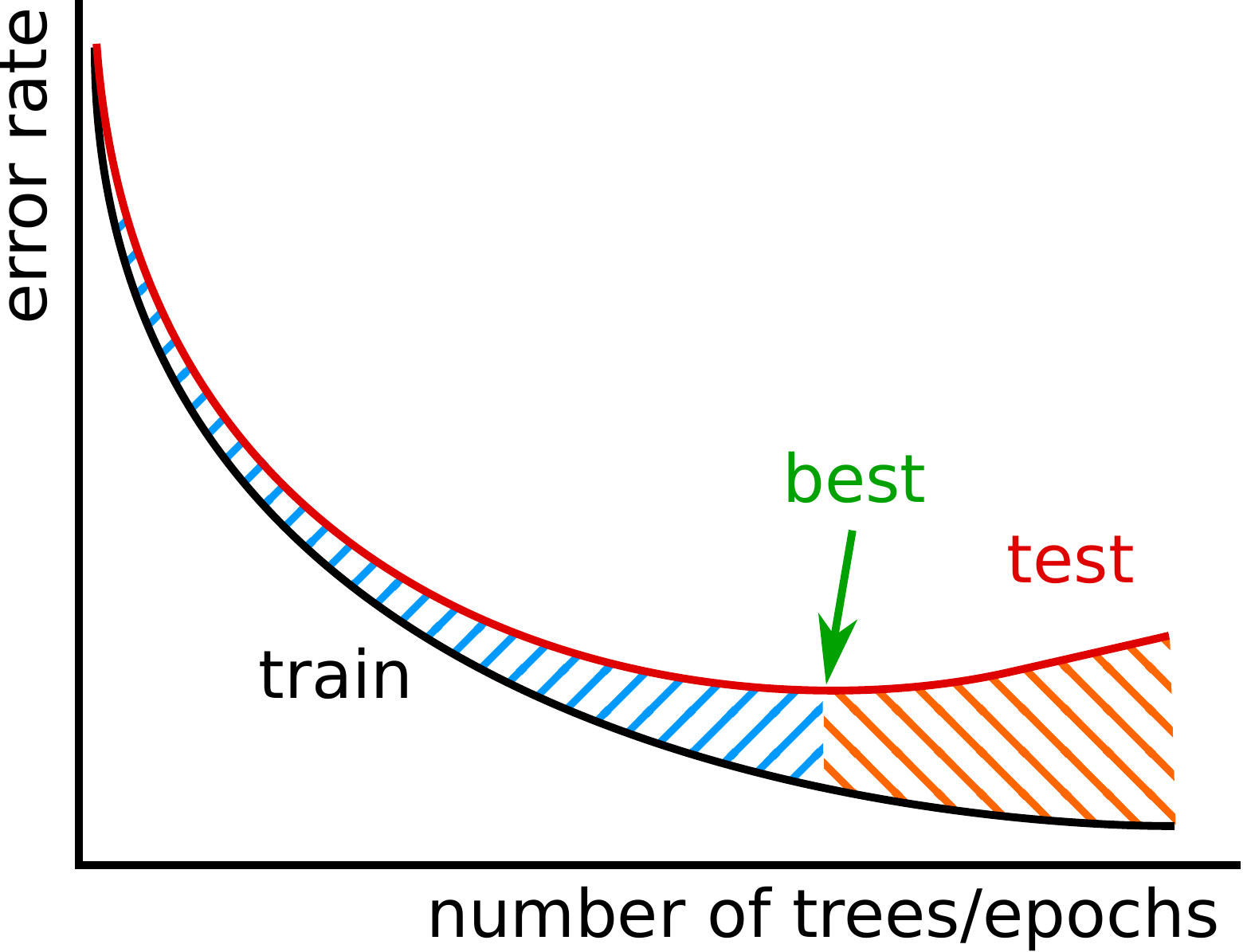}\label{fig:overtraining:large}}
  }
  \centerline{
    \subfigure[]
    {\includegraphics[width=.45\textwidth]{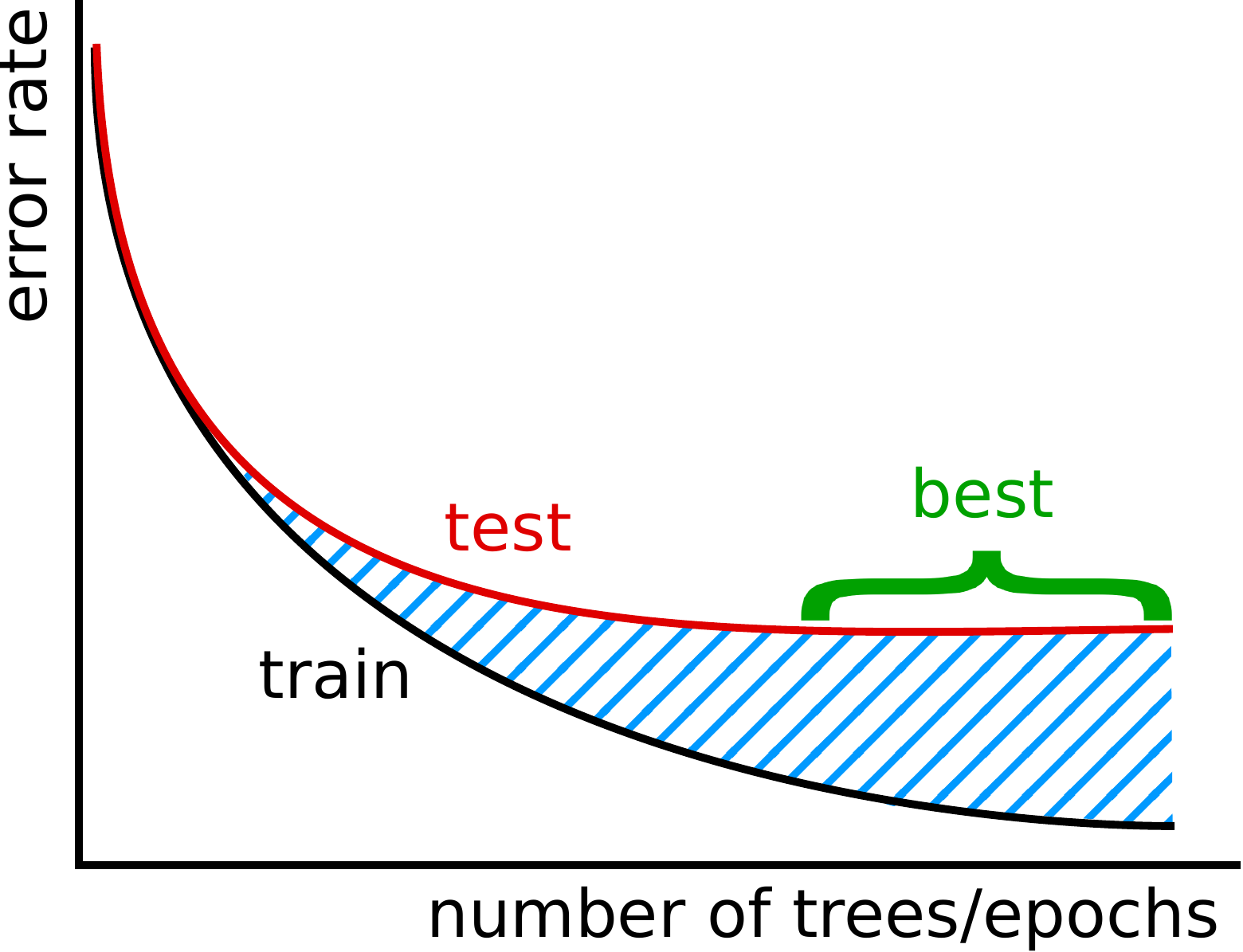}\label{fig:overtraining:flat}}
    \hspace*{10pt}
    \subfigure[]
    {\includegraphics[width=.45\textwidth]{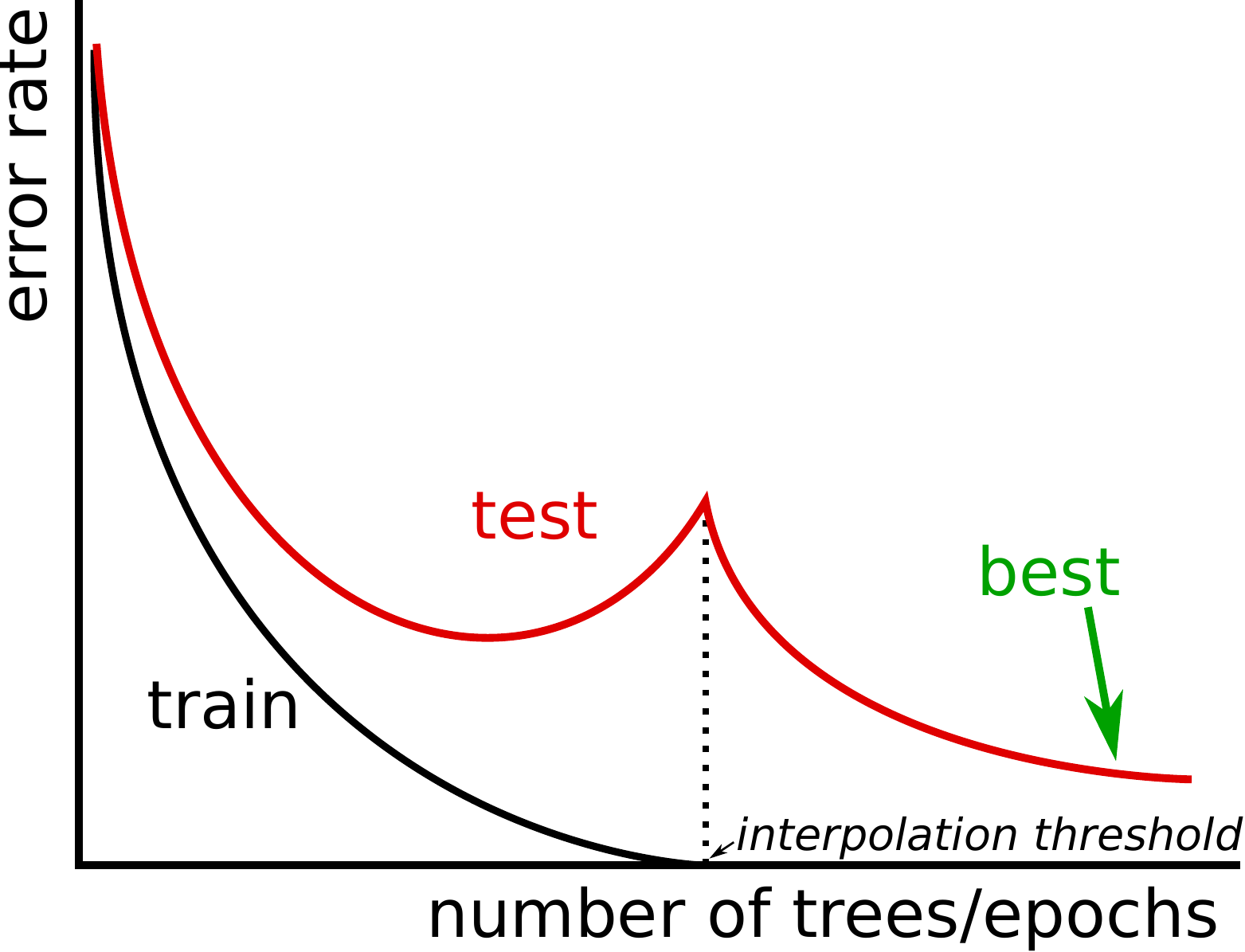}\label{fig:overtraining:interpolation}}
  }
  \caption{Overtraining estimation using the error rate as a function
    of the number of trees (for \bdts) or epochs (for neural
    networks). Black curves are measured on the training sample and
    red curves on the validation sample. The optimal
    classifier corresponds to the `best' label. The hatched areas
    represent overtraining: beneficial in blue (but underfitting), detrimental in orange (overfitting).
    (a) Typical curves, with the best model at the minimum of the
    testing curve, and overfitting beyond with decrease of
    performance. (b) The best model is overtrained but still improves
    performance. (c) Typical curves for \bdts with flattening testing
    error rate: all models in the flat area perform equally well
    despite increasing overtraining. (d) Interpolation regime: the
    best classifier is obtained after the training error has reached
    zero.}
  \label{fig:overtraining}
\end{figure}

It is important to check whether the model suffers from
overtraining. As shown in \fref{fig:overtraining} this can be achieved
by monitoring the error rate (or the loss function) during training,
as a function of the number of trees with \bdts
or training epoch with neural networks, on the training and validation
samples. \Fref{fig:overtraining:usual} is the canonical example of
such curves. The training error tends towards zero, while the testing
curve first follows the training curve, reaches a minimum and
increases again. The best classifier is the one at the minimum,
training further will reduce performance and cause overfitting: the
classifier has too much capacity (complexity) with respect to the
training sample. Selecting the model at the minimum means early
stopping~\cite{hastie2001elements}.

In many cases though, the situation is similar to
\fref{fig:overtraining:large}: the training and testing curves follow
each other but start diverging while still both improving. The
classifier is therefore already learning specificities of the training set, but
still learning properties that generalise well and improve performance
on the validation set. The testing curve goes through a minimum,
corresponding to the best model, and increases again, this time
showing detrimental overtraining as the performance decreases on the
validation set (overfitting regime). This is the typical U-shaped curve arising from the
bias--variance trade-off.

The curves could also look like \fref{fig:overtraining:flat}, where
the testing curve never goes through a minimum and instead
flattens. Once in the plateau, all classifiers are equivalent in terms
of performance on the validation set, while the training error keeps
improving (and could reach zero, this is the so-called interpolation
regime~\cite{Interpolation})). This is a typical curve for \bdts.

Finally the situation could correspond to
\fref{fig:overtraining:interpolation}. At the interpolation threshold
the training error reaches zero, but continued training of high
capacity classifiers leads to a double descent curve: the testing
performance keeps increasing while the training error stays at
zero~\cite{Belkin2019}.

\section{\DTs}
\label{sec:DT}

\DTs\ are a machine learning technique first developed in the context of data mining and
pattern recognition~\cite{Breiman}, which then gained momentum in various fields, including
medical diagnosis~\cite{Kononenko2001,Podgorelec2002}, insurance and loan screening, or optical
character recognition of handwritten text~\cite{Breiman}.

It was developed and formalised by Breiman \textit{et
  al.}~\cite{Breiman} who proposed the CART algorithm (Classification
And Regression Trees) with a complete and functional implementation of
\dts.

The basic principle is rather simple: it consists in extending a
simple cut-based analysis into a multivariate technique by continuing
to analyse events that fail a particular criterion. Many, if not most,
events do not have all characteristics of either signal or background
(for a two-class problem). The
concept of a \dt\ is therefore to not reject right away events that
fail a criterion, and instead to check whether other criteria may help
to classify these events properly.

In principle a \dt\ can deal with multiple output classes, each branch
splitting in many subbranches. In this chapter almost only binary
trees will be considered, with only two possible classes: signal and
background. The same concepts generalise to non-binary trees, possibly
with multiple outputs.

\Sref{sec:algoDT} describes the \dt building algorithm, controlled by
hyperparameters presented in \sref{sec:treeparam}. The way to split
nodes is explained in \sref{sec:split}, while \sref{sec:vars}
describes how \dts can advantageously deal with input variables and
how to optimise their list. Finally \sref{sec:Limitations} reports
several shortcomings of \dts, with suggestions to address them.

\subsection{Algorithm}
\label{sec:algoDT}
\begin{figure}
  \centerline{\includegraphics[width=.45\textwidth]{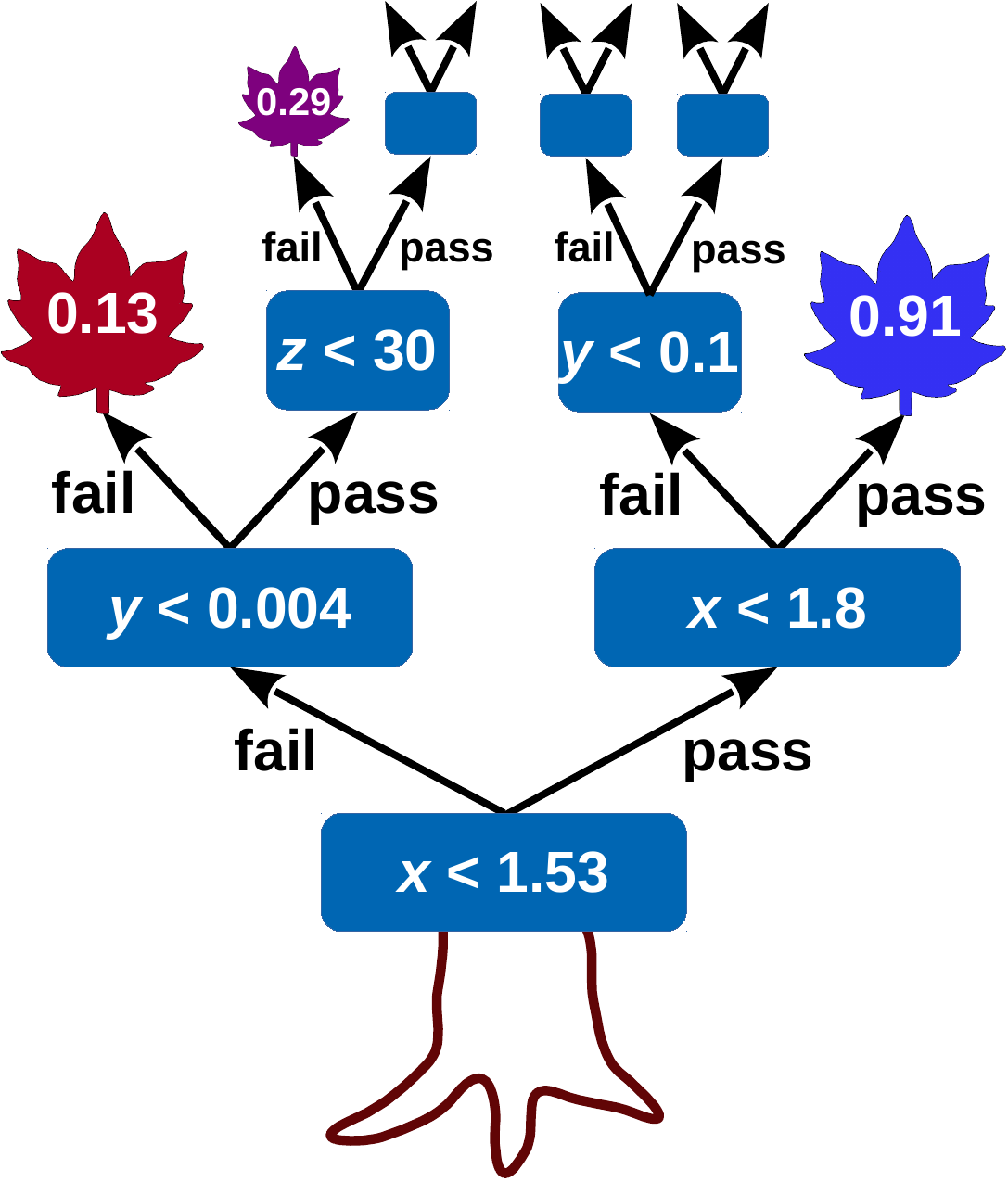}}
  \caption{Graphical representation of a \dt.  Blue rectangles
    are internal nodes with their associated splitting criterion;
    leaves are terminal nodes with their purity.}
  \label{fig:DT}
\end{figure}

Mathematically, \dts\ are rooted binary trees (as only trees with two
classes, signal and background, are considered). An example is shown
in \fref{fig:DT}. A \dt\ starts from an initial node, the root
node. Each node can be recursively split into two daughters or
branches, until some stopping condition is reached. The different
aspects of the process leading to a full tree, indifferently referred
to as growing, training, building or learning, are described in the
following sections.

Consider a sample of signal ($s_i$) and background ($b_j$) events,
each with weights $w_i^s$ and $w_j^b$, respectively, described by a set
$\vec x_i$ of variables. This sample constitutes the root node of a
new \dt.

Starting from this root node, the algorithm proceeds as follows:
\begin{enumerate}
\item If the node satisfies any stopping criterion, declare it as
  terminal (that is, a leaf) and exit the algorithm.
\item Sort all events according to each variable in $\vec x$.
\item For each variable, find the splitting value that gives the best
  separation between two children, one with mostly signal events, the
  other with mostly background events (see \sref{sec:split} for
  details). If the separation cannot be improved by any splitting,
  turn the node into a leaf and exit the algorithm.
\item Select the variable and splitting value leading to the best
  separation and split the node in two new nodes (branches), one
  containing events that fail the criterion and one with events that
  satisfy it.
\item Apply recursively from step 1 on each node.
\end{enumerate}
This is a greedy algorithm, not guaranteed to find the optimal solution.
At each node, all variables can be considered, even if they have been
used in a previous iteration: this allows to find intervals of
interest in a particular variable, instead of limiting oneself to
using each variable only once.

It should be noted that a \dt\ is human readable:
exactly which criteria an event satisfied in order to reach a
particular leaf can be traced. It is therefore possible to interpret a tree in
terms of, e.g., physics, defining selection rules, rather than only as
a mathematical object.

In order to make the whole procedure clearer, let us take the tree in
\fref{fig:DT} as an example. Consider that all events are
described by three variables: $x$, $y$ and $z$. All signal and
background events make up the root node.

All events are first sorted according to each variable:
\[\begin{array}{l}
 x^{s_{1}}\le x^{b_{34}}\le \cdots \le x^{b_{2}} \le x^{s_{12}},\\
 y^{b_{5}}\le y^{b_{3}}\le \cdots \le y^{s_{67}} \le y^{s_{43}},\\
 z^{b_{6}}\le z^{s_{8}}\le \cdots \le z^{s_{12}} \le z^{b_{9}},
\end{array}
\]
where superscript $s_i$ ($b_j$) represents signal (background) event
$i$ ($j$). Using some measure of separation between classes (see below) the
best splitting for each variable may be (arbitrary unit):
\[\begin{array}{lcl}
 x<1.53 & &\text{separation} = 5,\\
 y<0.01 & &\text{separation} = 3,\\
 z<25 & &\text{separation} = 0.7.
  \end{array}
\]

The best split is $x<1.53$, and two new
nodes are created, the left one with events failing this criterion and the right
one with events satisfying it. The same algorithm is applied
recursively to each of these new nodes. As an example consider the
right-hand-side node with events that satisfied $x<1.53$. After
sorting again all events in this node according to each of the three
variables, it was found that the best criterion was $x<1.8$, and
events were split accordingly into two new nodes. This time the
right-hand-side node satisfied one of the stopping conditions and was
turned into a leaf. From signal and background training events in this
leaf, the purity was computed as $p=0.91$. The left-hand-side node keeps splitting further.

The \dt\ output for a particular event $i$ is defined by how its
$\vec x_i$ variables behave in the tree:
\begin{enumerate}
\item Starting from the root node, apply the first criterion on $\vec
  x_i$.
\item Move to the passing or failing branch depending on the result of
  the test.
\item Apply the test associated to this node and move left or right in
  the tree depending on the result of the test.
\item Repeat step 3 until the event ends up in a leaf.
\item The \dt\ output for event $i$ is the value associated with this
  leaf.
\end{enumerate}

There are several conventions used for the value attached to a leaf.
It can be the purity $p=\frac{s}{s+b}$ where $s$ ($b$) is the sum of
weights of signal (background) events that ended up in this leaf
during training. It is then bound to $[0,1]$, close to 1 for signal
and close to 0 for background.

It can also be a binary answer, signal or background (mathematically
typically $+1$ for signal and 0 or $-1$ for background) depending on
whether the purity is above or below a specified critical value (e.g.
$+1$ if $p>\frac{1}{2}$ and $-1$ otherwise).

Looking again at the tree in \fref{fig:DT}, the leaf with purity
$p=0.91$ would give an output of 0.91, or $+1$ as signal if choosing a
binary answer with a critical purity of 0.5.

\subsection{Tree hyperparameters}
\label{sec:treeparam}
The number of hyperparameters of a \dt\ is relatively limited. The first
one is not specific to \dts\ and applies to most techniques requiring
training: how to normalise signal and background with respect to each other before starting the
training? Conventionally the sums of weights of signal and background
events are chosen to be equal (balanced classes), giving the root node a purity of 0.5,
that is, an equal mix of signal and background. \DTs are not
particularly sensitive to this original normalisation as in practice,
a few early splits will produce nodes with more balanced categories,
therefore only leading to a limited inefficiency in the training
process which only impacts marginally the final discriminating
power.

Other hyperparameters concern the selection of splits. A
list of discriminating variables is needed, and a way to evaluate the
best separation between signal and background events (the goodness of
the split). Both aspects are described in more detail in
\sref{sec:split} and \sref{sec:vars}.

The splitting has to stop at some point, declaring such nodes as
terminal leaves. Conditions to satisfy can include:
\begin{itemize}
\item a minimum leaf size. A simple way is to require at least
  $N_\text{min}$ training events in each node after splitting, to
  ensure the statistical significance of the purity measurement, with a
  statistical uncertainty $\sqrt{N_\text{min}}$. It becomes a little bit
  more complicated with weighted events, as is normally the case in
  high-energy physics applications.
  Using the effective number of events instead may be considered:
  \[N_\text{eff}=\frac{\big(\sum_{i=1}^{N}w_i\big)^2}{\sum_{i=1}^{N}w_i^2},\]
  for a node with $N$ events associated to weights $w_i$ ($N_\text{eff}=N$
  for unweighted events). 
\item having reached perfect separation (all events in the node belong
  to the same class).
\item an insufficient improvement with further splitting.
\item a maximum tree depth, if the tree
  cannot have more than a certain number of layers (for purely
  computational reasons or to have like-size trees).
\end{itemize}

Finally a terminal leaf has to be assigned to a class. This is
classically done by labelling the leaf as signal if $p>0.5$ and
background otherwise.

\subsection{Splitting a node}
\label{sec:split}
The core of a \dt\ algorithm resides in how a node is split into two.
Consider an impurity measure $i(t)$ for node $t$, which describes to
what extent the node is a mix of signal and background. Desirable
features of such a function are that it should be:
\begin{itemize}
\item maximal for an equal mix of signal and background (no separation).
\item minimal for nodes with either only signal or only background events
  (perfect separation).
\item symmetric in signal and background purities, as isolating
  background is as valuable as isolating signal.
\item strictly concave in order to reward purer nodes. This tends to
  favour asymmetric end cuts with one smaller node and one larger node.
\end{itemize}

A figure of merit can be constructed with this impurity function, as
the decrease of impurity for a split $S$ of node $t$ into two children
$t_P$ (pass) and $t_F$ (fail):
\[\Delta i(S,t)=i(t)- p_P\cdot i(t_P) - p_F\cdot i(t_F),\]
where $p_P$ ($p_F$) is the fraction of events that passed (failed)
split $S$.

The goal is to find the split $S^*$ that maximises the decrease of
impurity:
\[\Delta i(S^*,t) = \max_{S\in \{\text{splits}\}}\Delta i(S,t).\]
It will result in the smallest residual impurity, which minimises the
overall tree impurity.

A stopping condition can be defined using the decrease of impurity,
not splitting a node if $\Delta i(S^*,t)$ is less than
some predefined value. Such early-stopping criterion requires care, as sometimes a seemingly very weak
split may allow child nodes to be powerfully split further (see
\sref{sec:prune} about pruning).

\begin{figure}
  \centerline{\includegraphics[width=.5\textwidth]{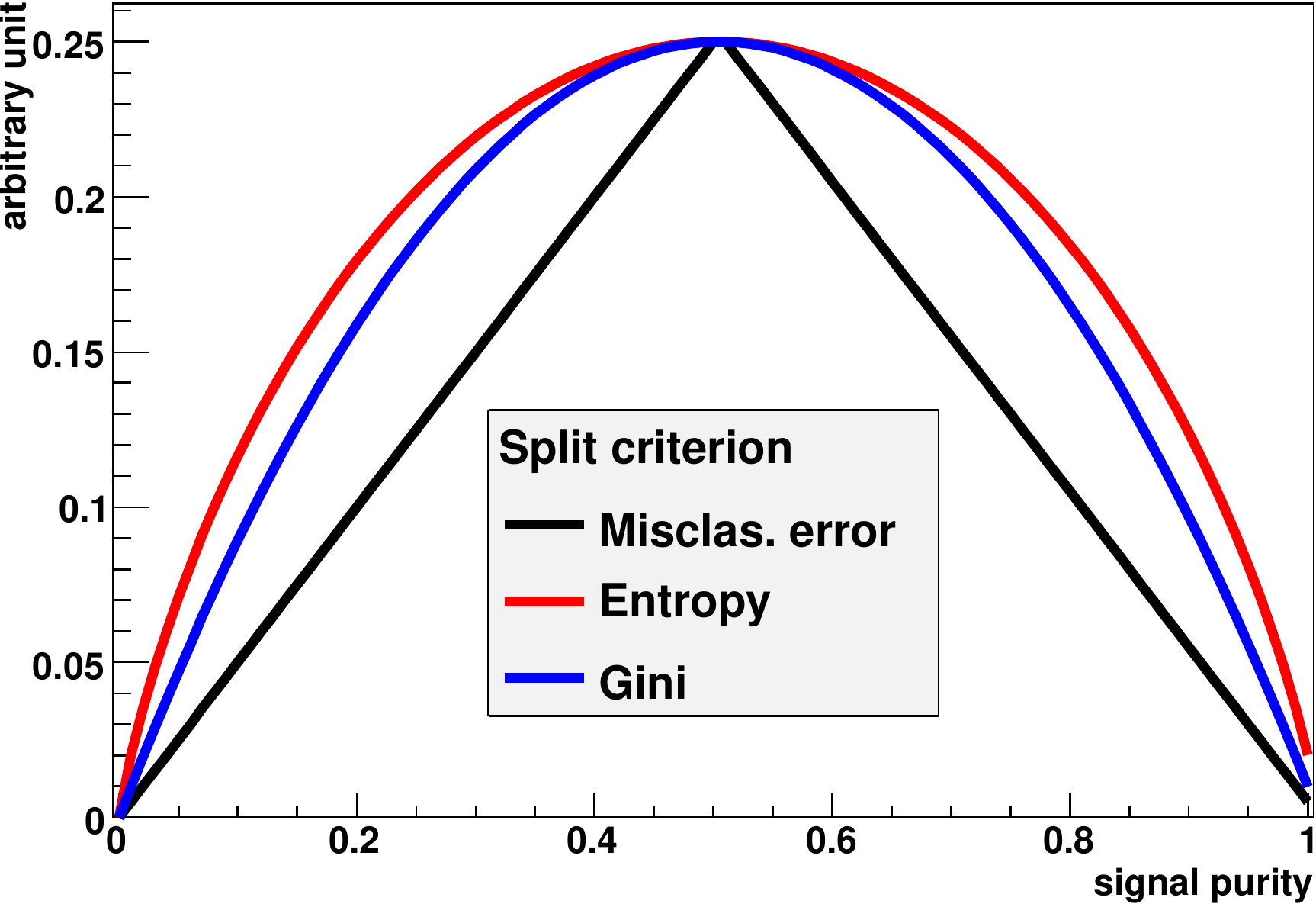}}
  \caption{Impurity measures as a function of signal purity.}
  \label{fig:concave}
\end{figure}

Common impurity functions
(exhibiting most of the desired features mentioned previously) are
illustrated in \fref{fig:concave}:
\begin{itemize}
\item the misclassification error: $1-\max(p,1-p)$,
\item the (cross) entropy~\cite{Breiman}: $-\sum_{i=s,b} p_i\log p_i$, with $p_b = 1-p_s$ and $p_s = p$,
\item the Gini index of diversity~\cite{Gini}.
\end{itemize}
The Gini index is the most popular in \dt\ implementations. It
typically leads to similar performance to entropy.

Other measures are also used sometimes, which do not satisfy all
criteria listed previously but attempt at optimising signal
significance, a typical final goal in high-energy physics applications
(see \sref{sec:significance}):
\begin{itemize}
\item cross section significance (optimising $\frac{s}{\sqrt{s+b}}$): $-\frac{s^2}{s+b}$,
\item excess significance (optimising $\frac{s}{\sqrt{b}}$): $-\frac{s^2}{b}$.
\end{itemize}

\subsection{Variable selection}
\label{sec:vars}
Overall \dts\ are very resilient to most factors affecting
variables. They are not too much affected by the `curse of
dimensionality', which forbids the use of too many variables in most
multivariate techniques. For \dts\ the CPU consumption scales as
$nN\log N$ with $n$ variables and $N$ training events. It is not
uncommon to encounter \dts\ using tens~\cite{D02} or hundreds~\cite{Miniboone2} of variables,
although this is usually frowned upon in high-energy physics: more
variables means more distributions and correlations to check, more
complex interplay with systematic uncertainties, more dependence on
the Monte Carlo event properties that are usually used during training
and may not match real data so well, so physicists tend to reduce the list of discriminating variables to typically 10--15. On the other hand adding variables tends to always improve the
performance of \dts (see \sref{sec:UseCases} for an example).

\subsubsection{Manipulating variables}
With most machine learning algorithms, a careful preparation of inputs
is necessary to achieve good performance. Although not detrimental to
\dts, such manipulations are not really compulsory as \dts tend to be
very stable under such transforms.

A \dt\ is immune to duplicate variables: the sorting of events
according to each of them would be identical, leading to the exact
same tree. The order in which variables are presented is completely
irrelevant: all variables are treated equal. The order of events in
the training samples is also irrelevant.

If variables are not very discriminating, they will simply be ignored and
will not add any noise to the \dt. The final performance will not be
affected, it will only come with some CPU overhead during both
training and evaluation.

\DTs\ can deal easily with both continuous and discrete variables,
simultaneously.

Another typical task before training a multivariate technique is to
transform input variables by for instance making them fit in the same
range (normalisation), having unit variance (standardisation) or
taking the logarithm to regularise the variable. This is totally
unnecessary with \dts, which are completely insensitive to the
replacement of any subset of input variables by (possibly different)
arbitrary strictly monotone functions of them (e.g. converting MeV to GeV),
as the same ordering of events would induce the same splits on the dataset, producing the same \dt.  This means that \dts\ have some
immunity against outliers. The above is strictly true only if testing
all possible cut values while evaluating the optimal split. If there
is some computational optimisation (e.g., check only 20 possible cuts
on each variable), it may not work anymore and some transformation of
inputs may be beneficial, at the very least to speed up convergence
(numerical precision could also be a factor).

If linear correlations exist between variables, first decorrelating the input
variables and then feeding them to the \dt may help. If not doing this
decorrelation, a \dt\ will anyway find the correlations but in a very
suboptimal way, by successive approximations, adding complexity to the
tree structure without performance gain.

\subsubsection{Mean decrease impurity}
\label{sec:mdi}
It is possible to rank variables in a \dt, adding up the decrease of impurity (see \sref{sec:split}) for
each node where the variable was used to split, hence computing the
mean decrease impurity (MDI). The variable with the largest
decrease of impurity is the best variable. A shortcoming of this
approach is that it is computed on the training set only, and may be
exaggerating the importance of some variables because of overfitting.

There is another shortcoming with variable ranking in a \dt: variable
masking. Variable $x_j$ may be just a little worse than variable $x_i$
and would end up never being picked in the \dt\ growing
process. Variable $x_j$ would then be ranked as irrelevant.
But if $x_i$ were removed, then $x_j$ would become very relevant. Note that
this is not important in terms of pure performance of the tree: it did
find the optimal way to use both variables in this particular
training. If trying to learn something from
the tree structure on the other hand, like deriving selection rules, this phenomenon will interfere with the potential understanding.

There is a solution to this feature, called surrogate
splits~\cite{Breiman}. For each split, a comparison is made between training
events that pass or fail the optimal split and events that pass or fail a
split on another variable. The split that mimics best the optimal
split is called the surrogate split. This can be taken into
consideration when ranking variables. It has applications in case of
missing data: the optimal split can be replaced by the surrogate
split.

All in all, variable rankings should never be taken at face
value. They do provide valuable information but should not be
over-interpreted.

\subsubsection{Permutation importance}
\label{sec:mda}
The shortcomings of MDI discussed above are partially addressed with a
different technique called permutation importance or mean decrease
accuracy (MDA). While MDI mostly works for \dts, permutation
importance is suited for all models using tabular data. It is defined
as the decrease of performance of an already trained model when
applying it on a sample after randomly shuffling a single
discriminating variable~\cite{RandomForests}. If the variable is of
any use, the performance should decrease when submitted to this noisy
input, and more so if the tree relies heavily on this feature for its
prediction. Repeating this for all input variables, the importance of each of them can be ranked. The operation can be done
multiple times, shuffling each variable differently, in order to get a
mean value and uncertainty on variable importance. As with MDI, the
measured importance is not telling anything about the intrinsic merit
of a single variable (in terms of physics meaning for instance), but
is rather a measure of its importance for this particular training.

Another advantage of this approach is that it can be applied on the
validation set as well. Variables that are important on the training
set but not on the validation set may be a source of overfitting.

As with MDI however, correlations may hide the intrinsic performance
of a variable. If two variables are correlated and only one is
shuffled, the proper information is still accessible, giving a lower
importance to both. Once again, interpreting variable rankings must be
done with care.

\subsubsection{Choosing variables}
It may sound obvious that only well discriminating
variables should be used as input features to the \dt training. It is nevertheless
not trivial to achieve: variables are often correlated, they come in large
numbers, and can be more or less discriminating in various regions of
the input-feature phase space. The \dt will isolate sub-regions, whose
properties are not readily available when measuring any kind of
discrimination in the full training set.

Brute force is a possibility: with a limited number of $N$ features,
train all possible combinations of $N$, $N-1$, etc.,
variables, and pick the best one according to some metric (see
\sref{sec:fom}). In reality this becomes quickly impractical.

Instead, a commonly used approach in high-energy physics is backward elimination~\cite{UnderstandingML}, which starts from the full list of
$N$ variables used to train a tree ($T_N$). Then train all \dts with $N-1$
variables and keep the best performing one on the validation set
($T_{N-1}$). Starting from these $N-1$ variables, train all \dts with $N-2$
variables to build $T_{N-2}$, and so on. Usually the performance of
tree $T_k$ will decrease with $k$, and it is up to the analyser to
decide how much performance to lose compared to getting a simpler
(possibly more robust) tree. This is the usual trade-off of cost and
complexity.

The selection can also be done in reverse, starting from $k=1$
variable, training all trees with $k+1$ variables, keeping the best
one on the validation set and moving to $k+2$ variables, until
$k=N$ (forward greedy selection~\cite{UnderstandingML}). The advantage is that one can stop adding variables once the
performance curve seems to saturate. It is on the other hand not
equivalent to backward elimination, as it may miss powerful variable
combinations.

It can be tempting to train a tree with many variables and then remove
the lowest ranked. Although quicker, it will most certainly be
suboptimal because of the shortcomings of such rankings, as described
in \sref{sec:mdi} and \sref{sec:mda}. The ranking is only
relevant to the corresponding tree, and as soon as one of the
variables is removed the others may be reshuffled.

\subsection{Limitations}
\label{sec:Limitations}

Despite all the nice features presented above, \dts\ are known to be
relatively unstable. If trees are too optimised for the training
sample, they may not generalise very well to unknown events, as they
would depend on the training sample (see
\sref{sec:samples}). This can be mitigated with pruning,
described in \sref{sec:prune}. Combining several classifiers
can also improve the overall performance, as shown in
\sref{sec:average}.

\subsubsection{Training sample composition}
\label{sec:samples}
A small change in the training sample can lead to drastically
different tree structures (high variance), rendering the physics interpretation a bit
less straightforward. As such, a \dt is not stable, where stability
means that a slight change of the inputs does not change much the
output~\cite{UnderstandingML}. For sufficiently large training samples, the
performance of these different trees will be equivalent, but on small
training samples variations can be very large. This does not give too
much confidence in the result.

Moreover a \dt\ output is by nature discrete, limited by the purities
of all leaves in the tree. To decrease the
discontinuities the tree size and complexity has to increase,
which may not be desirable or even possible. Then the tendency is to
have spikes in the output distribution at specific purity values, or
even two delta functions at $\pm 1$ if using a binary answer rather
than the purity output.

\subsubsection{Pruning a tree}
\label{sec:prune}
When growing a tree, each node contains fewer and fewer events,
leading to an increase of the statistical uncertainty on each new
split. The tree will tend to become more and more
specialised, focusing on properties of the training sample that may
not reflect the expected result, had there been infinite statistics to
train on. Its variance increases.

A first approach to mitigate this effect and keep the variance under control, sometimes referred to as
pre-pruning, has already been described in \sref{sec:DT}, using
stopping conditions.
The limitation is that requiring too big a minimum leaf size or too much
of an improvement may prevent further splitting that could be very
beneficial later on.

Another approach consists in building a very large tree and then
cutting irrelevant branches (which target too closely the training
sample and would not generalise well) by turning an internal node and
all its descendants into a leaf, removing the corresponding
subtree. This is post-pruning, or simply pruning.

There are many different pruning algorithms available.
Expected error pruning~\cite{Quinlan} starts from a fully grown tree
and compares the expected error of a node to the weighted sum of
expected errors from its children. If the expected error of the node
is less than that of the children, then the node is pruned. This does
not require a separate pruning sample. With reduced error
pruning~\cite{Quinlan} the misclassification rate on a pruning sample
for the full tree is compared to the misclassification rate when a
node is turned into a leaf. If the simplified tree has better
performance, the subtree is pruned. Finally cost--complexity pruning is
part of the CART algorithm~\cite{Breiman} and the most used. Starting
from a fully grown tree, the cost--complexity is computed as the sum of
misclassification rate and a term proportional to the number of nodes
in the tree (the complexity part, penalising larger trees).  A
sequence of decreasing cost--complexity subtrees is generated, and
their misclassification rate on the pruning sample is computed. It
will first decrease, and then go through a minimum before increasing
again. The optimally pruned tree is the one corresponding to the
minimum.

It should be noted that the best pruned tree may not be optimal or
necessary when part of a forest of trees, such as those introduced in
the next Sections.

\subsubsection{Ensemble learning}
\label{sec:average}
Pruning is helpful in maximising the generalisation
potential of a single \dt. It nevertheless does not address other
shortcomings of trees like the discrete output or lack of stability. A way out is to
proceed with averaging several trees, with the added potential bonus
that the discriminating power may increase. Such approaches belong to
the general theoretical framework of ensemble
learning~\cite{Friedman2008}. Many
averaging techniques have been developed. Bagging, boosting and random forests are such
techniques and will be described in the following Sections.

\begin{figure}
  \centerline{
    \subfigure[]
    {\includegraphics[height=2.8cm]{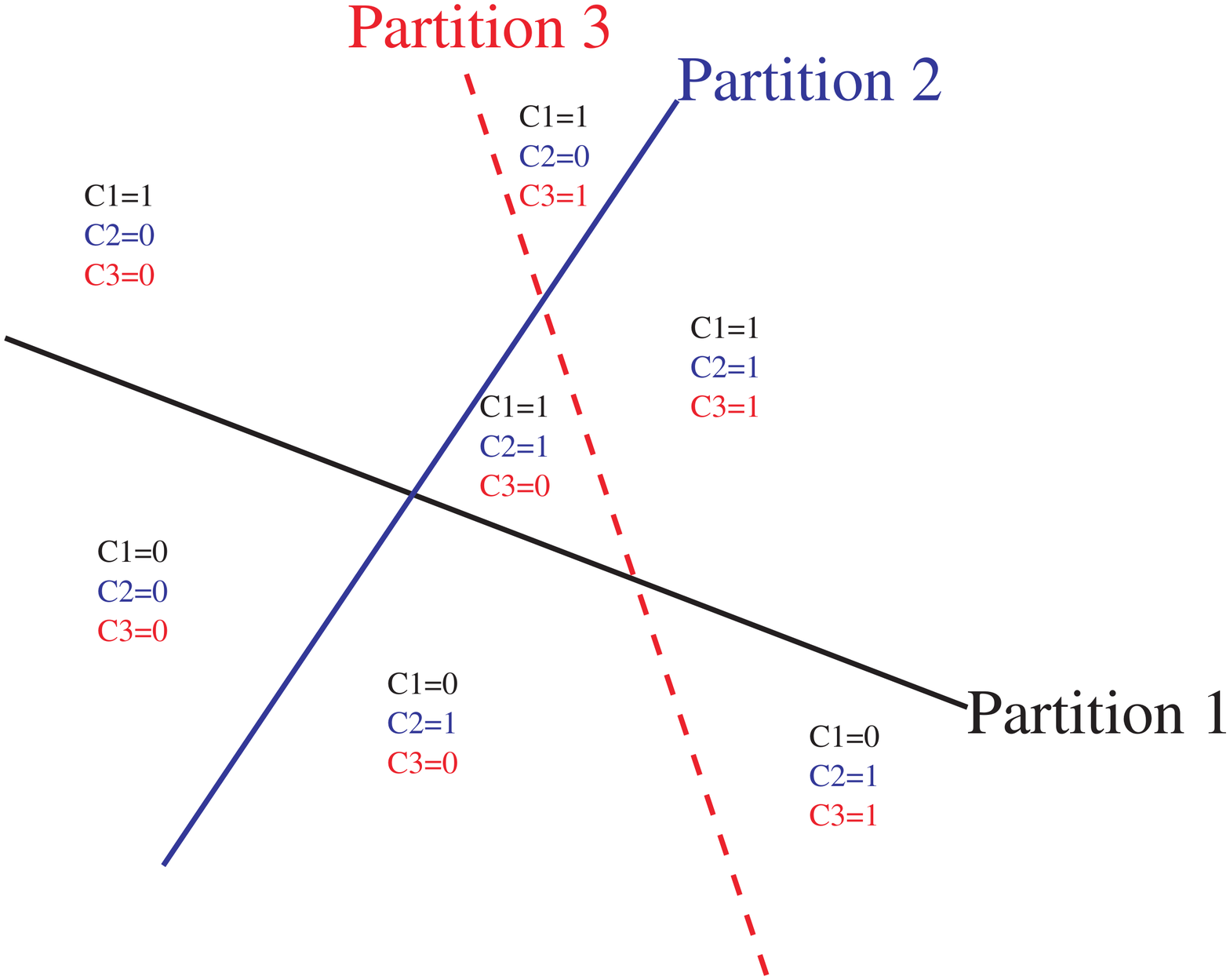}\label{fig:limitations:distributed}}
    \hspace*{10pt}
    \subfigure[]
    {\includegraphics[height=2.8cm]{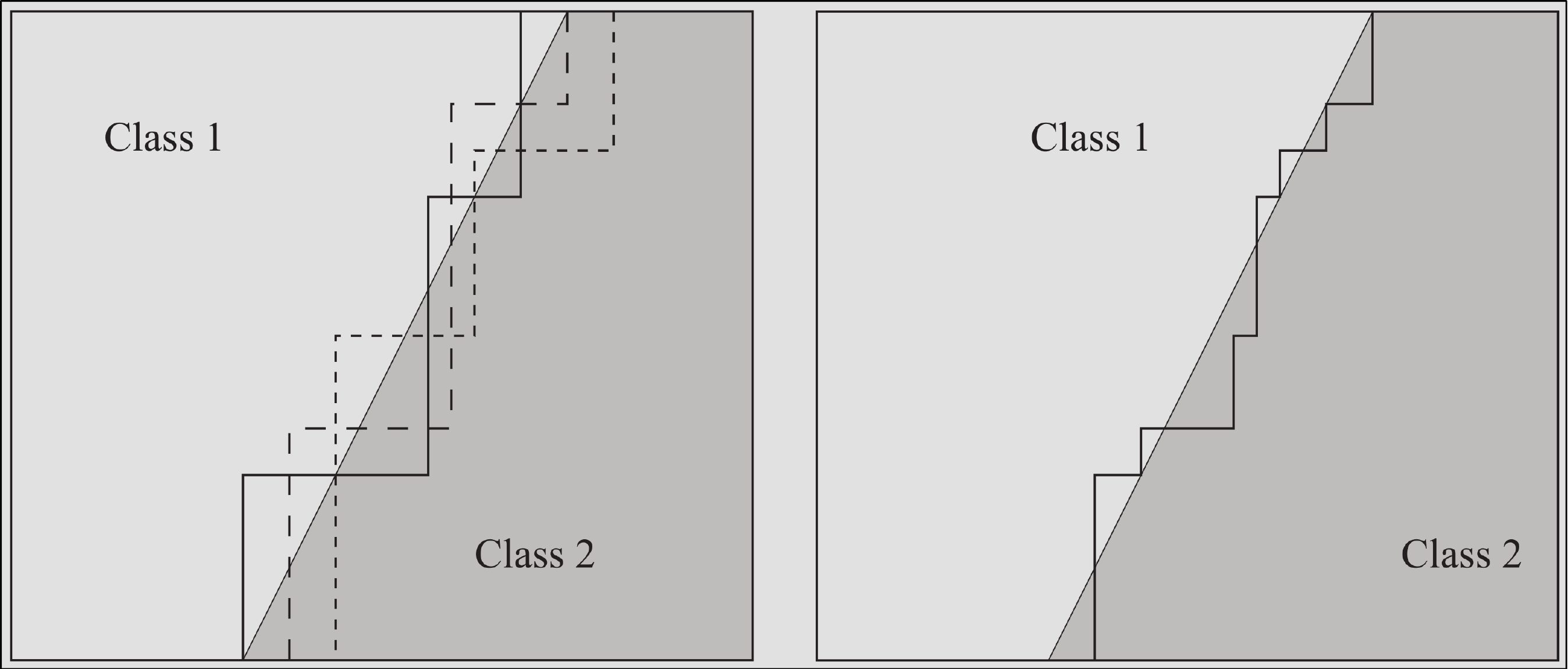}\label{fig:limitations:dtbdt}}
  }
  \caption{(a) Description of 2D space combining three discriminants. (b) Three separate decision trees and their combination~\cite{Dietterich97}.}
  \label{fig:limitations}
\end{figure}

The power of ensemble learning resides in the much richer description
of the input patterns when using several classifiers
simultaneously. It is applicable to other machine learning techniques
than \dts. As shown in the example of
\fref{fig:limitations:distributed} in a simple 2D case, a
classifier may split the space in two (partitions 1/2/3), but three
classifiers each doing this can possibly give more complete
information about seven regions, each region being represented by
three numbers (C1/C2/C3). When all three classifiers give the
same answer, the confidence increases. Using \dts as in
\fref{fig:limitations:dtbdt}, three simple \dts give a crude
separation of classes 1 and 2, while averaging them produces a
decision contour that is much closer to the actual class separation.

\section{\BDTs}
\label{sec:BDT}

As will be shown in this section, the boosting algorithm has turned
into a very successful way of improving the performance of any type of
classifier, not only decision trees. After a short history of boosting
in \sref{sec:history}, the generic algorithm is presented in
\sref{sec:boostalgo} and specific implementations (AdaBoost and
gradient boosting) are described in Secs.~\ref{sec:adaboost} and
\ref{sec:gradboost}. Boosting is illustrated with a few examples in
\sref{sec:examples}. Other boosting implementations are shown in
\sref{sec:otherboost}. The use of boosting for regression rather than
classification is presented in \sref{sec:regression}. Finally the
application of \bdts in high-energy physics, where it is so far the
machine learning algorithm of choice, is illustrated in
\sref{sec:hep}.

\subsection{Introduction}
\label{sec:history}
The first provable algorithm of boosting was proposed in
1990~\cite{Schapire}. It worked in the following way:
\begin{itemize}
\item train a classifier $T_1$ on a sample of $N$ events;
\item train $T_2$ on a new sample with $N$ events, half of which were
  misclassified by $T_1$;
\item build $T_3$ on events where $T_1$ and $T_2$ disagree.
\end{itemize}
The boosted classifier was defined as a majority vote on the outputs
of $T_1$, $T_2$ and $T_3$.

Following up on this idea boosting by majority~\cite{Freund} was
introduced in 1995. It consisted in combining many learners with a
fixed error rate. This was an impractical prerequisite for a viable
automated algorithm, but was a stepping stone to the first functional
boosting algorithm, called AdaBoost~\cite{Adaboost}.

Boosting, and in particular \bdts, have become increasingly
popular in high-energy physics and are extensively used in physics
analyses and object identification at the Tevatron and the LHC (see
\sref{sec:hep} for a few examples).

\subsection{Boosting algorithm}
\label{sec:boostalgo}
It is hard to make a very good
discriminant, but relatively easy to make simple ones which are
certainly more error-prone (high bias) but are still performing at least
marginally better than random guessing. Such discriminants are called
weak classifiers. The goal of boosting is to combine
such weak classifiers into a new, more stable one, with a smaller
error rate (with lower bias than the individual classifiers) and better performance.

Consider a training sample $\mathbb{T}_k$ containing $N_k$ events. The
$i^\text{th}$ event is associated with a weight $w_i^k$, a vector of
discriminating variables $\vec x_i$ and a class label $y_i=+1$ for signal,
$-1$ for background. The pseudocode for a generic boosting algorithm
is:

\verb+  Initialise+ $\mathbb{T}_1$ 

\verb+  for+ $k$ \verb+in 1..+$N_\text{tree}$ 

\verb+    train classifier+ $T_k$ \verb+on+ $\mathbb{T}_k$ 

\verb+    assign weight+ $\alpha_k$ \verb+to+ $T_k$ 

\verb+    modify+ $\mathbb{T}_k$ \verb+into+ $\mathbb{T}_{k+1}$

The boosted output is some function $F(T_1,..,T_{N_\text{tree}})$,
typically a weighted average:
\[F(i)=\sum_{k=1}^{N_\text{tree}}\alpha_k T_k(\vec x_i).\]

Thanks to this averaging, the output becomes quasi-continuous,
mitigating one of the limitations of single \dts (see
\sref{sec:samples}).

Note that in this process, once a particular tree is trained it is
never modified, but just added to the mix. This is a different
approach from, e.g., neural networks, in which the same weights are
repeatedly updated over epochs to converge towards the final classifier.

\subsection{AdaBoost}
\label{sec:adaboost}
One particularly successful implementation of the boosting algorithm
is AdaBoost~\cite{Adaboost}. AdaBoost
stands for adaptive boosting, referring to the fact that the learning
procedure adjusts itself to the training data in order to classify it
better. There are many variations for the actual
implementation, and it is the most common boosting algorithm. It
typically leads to better results than without boosting, up to the
Bayes limit as will be seen later.

An actual implementation of the AdaBoost algorithm works as follows.
After having built tree $T_k$, events in the
training sample $\mathbb{T}_k$ that are misclassified by $T_k$ should be checked, hence
defining the misclassification rate $R(T_k)$. In order to ease the
math, let us introduce some notations. Define $\mathbb{I}:
X\to\mathbb{I}(X)$ such that $\mathbb{I}(X)=1$ if statement $X$ is true, and 0
otherwise. A function can now be defined that tells whether an event
is misclassified by $T_k$. In the \dt\ output convention of returning only
\{$\pm 1$\} it gives:
\[\text{isMisclassified}_k(i) = \mathbb{I}\big(y_i\times T_k(i)\le 0\big),\]
while in the purity output convention (with a critical purity of 0.5)
it leads to:
\[\text{isMisclassified}_k(i) = \mathbb{I}\big(y_i\times (T_k(i)-0.5)\le 0\big).\]
The misclassification rate is now:
\[ R(T_k) = \varepsilon_k=\frac{\sum_{i=1}^{N_k}w_i^k\times
  \text{isMisclassified}_k(i)}{\sum_{i=1}^{N_k}w_i^k}. \]
This misclassification rate can be used to derive a weight associated
to tree $T_k$:
\[\alpha_k=\beta\times\ln\frac{1-\varepsilon_k}{\varepsilon_k},\]
where $\beta$ is a free parameter to adjust the strength of
boosting (set to one in the original algorithm). Similarly to the
naming convention of other machine learning algorithms, it can be seen
as a learning rate or shrinkage coefficient and drives how aggressive
boosting should be.

The core of the AdaBoost algorithm resides in the following step: each
event in $\mathbb{T}_k$ has its weight changed in order to create a
new sample $\mathbb{T}_{k+1}$ such that:
\[ w_i^k \to w_i^{k+1}=w_i^k\times e^{\alpha_k\cdot\text{isMisclassified}_k(i)}. \]

This means that properly classified events are unchanged from
$\mathbb{T}_{k}$ to $\mathbb{T}_{k+1}$, while misclassified events see
their weight increased by a factor $e^{\alpha_k}$. The next tree
$T_{k+1}$ is then trained on the $\mathbb{T}_{k+1}$ sample. This next
tree will therefore see a different sample composition with more
weight on previously misclassified events, and will therefore try
harder to classify properly difficult events that tree $T_k$ failed to
identify correctly, while leaving alone those events that previous
iterations can handle properly. The final AdaBoost result for event
$i$ is:
\[T(i) = \frac{1}{\sum_{k=1}^{N_\text{tree}}\alpha_k}\sum_{k=1}^{N_\text{tree}}\alpha_k T_k(i). \]

As an example, assume for simplicity the case $\beta=1$. A not-so-good
classifier, with a misclassification rate $\varepsilon=40$\% would have a
corresponding $\alpha = \ln\frac{1-0.4}{0.4}=0.4$. All misclassified
events would therefore get their weight multiplied by $e^{0.4}=1.5$,
and the next tree will have to work a bit harder on these events. Now
consider a good classifier with an error rate $\varepsilon=5$\% and
$\alpha = \ln\frac{1-0.05}{0.05}=2.9$. Misclassified events get a
boost of $e^{2.9}=19$ and will contribute decisively to the structure
of the next tree! This shows that being failed by a good classifier
brings a big penalty.

It can be shown~\cite{FSepsilon} that the misclassification rate
$\varepsilon$ of the boosted result on the training sample is bounded
from above:
\[\varepsilon \le \prod_{k=1}^{N_\text{tree}} 2\sqrt{\varepsilon_k(1-\varepsilon_k)}.\]
If each tree has $\varepsilon_k\ne 0.5$, that is to say, if it does
better than random guessing, then the conclusion is quite remarkable:
the error rate falls to zero for a sufficiently large
$N_\text{tree}$. A corollary is that the training data is overfit.

Overtraining is usually regarded as a negative feature. Does this
mean that boosted \dts\ are doomed because they are too powerful on
the training sample? Not really. As shown in \sref{sec:overtraining}
what matters most is not the error rate on the training sample, but
rather the error rate on the testing sample. In the case of
\fref{fig:overtraining:usual} or \fref{fig:overtraining:large} boosting
should stop when the minimum is reached (early stopping). It has however been
routinely observed~\cite{BoostingFoundations,friedman2001,Proba} that \bdts\ often do not go through such a minimum, but
rather tend towards a plateau in testing error
(see \fref{fig:overtraining:flat}). Boosting could be stopped
after having reached this plateau.

\begin{figure}
  \centerline{
    \subfigure[]
    {\includegraphics[width=.44\textwidth]{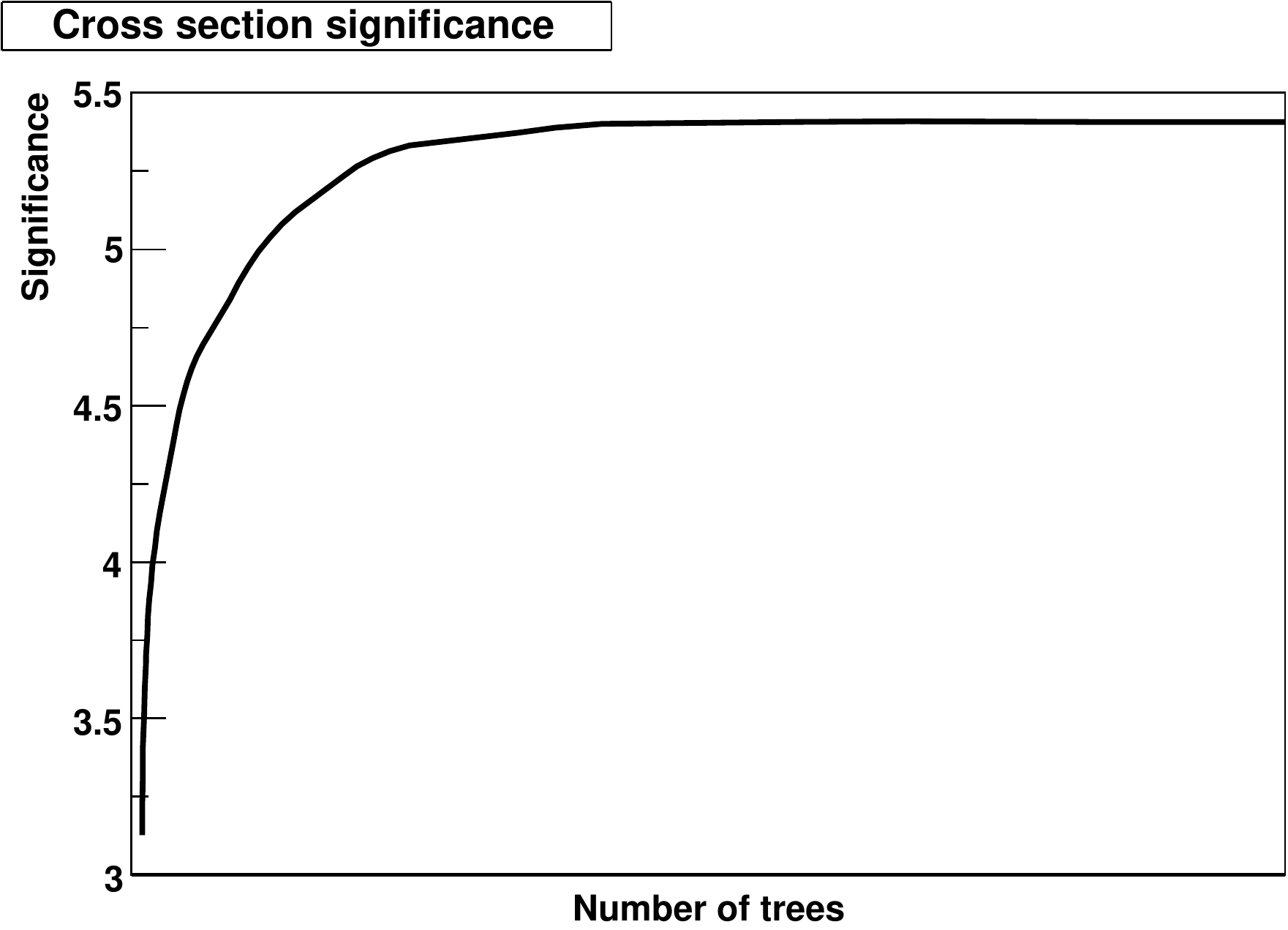}\label{fig:boost:sig}}
    \hspace*{10pt}
    \subfigure[]
    {\includegraphics[width=.44\textwidth,height=.35\textwidth]{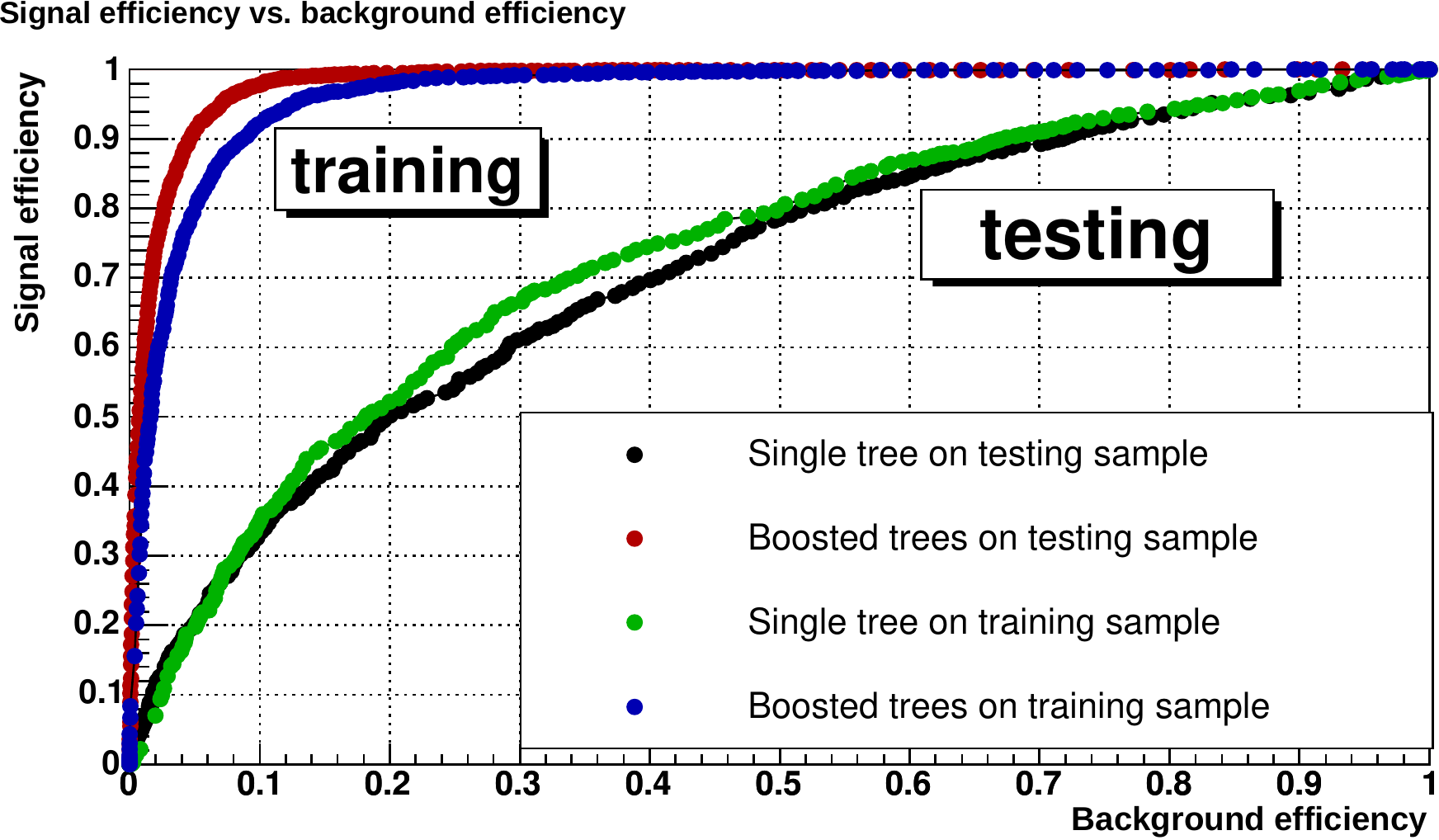}\label{fig:boost:traintest}}
    }
  \centerline{
    \subfigure[]
    {\includegraphics[width=.44\textwidth]{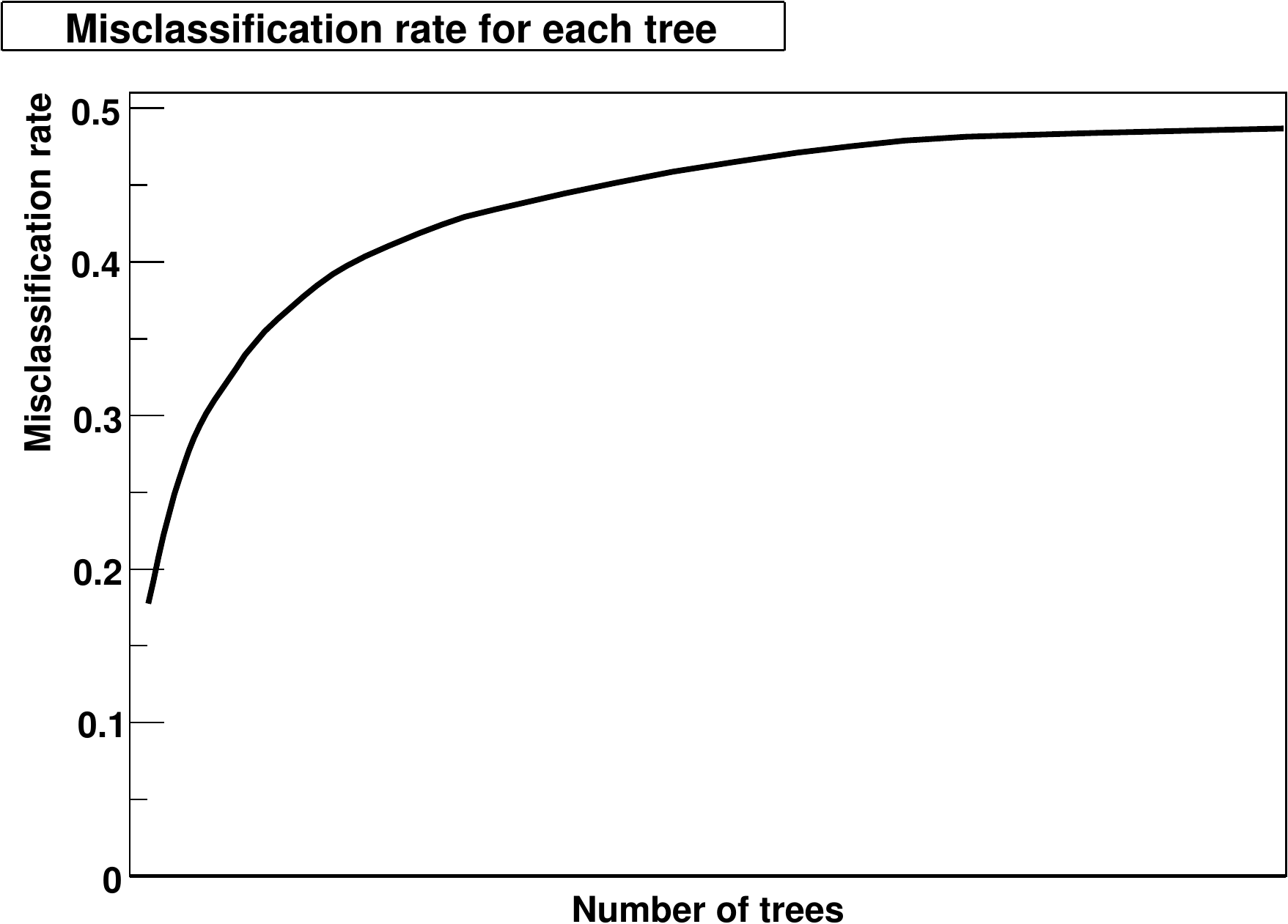}\label{fig:boost:errorrate}}
    \hspace*{10pt}
    \subfigure[]
    {\includegraphics[width=.44\textwidth]{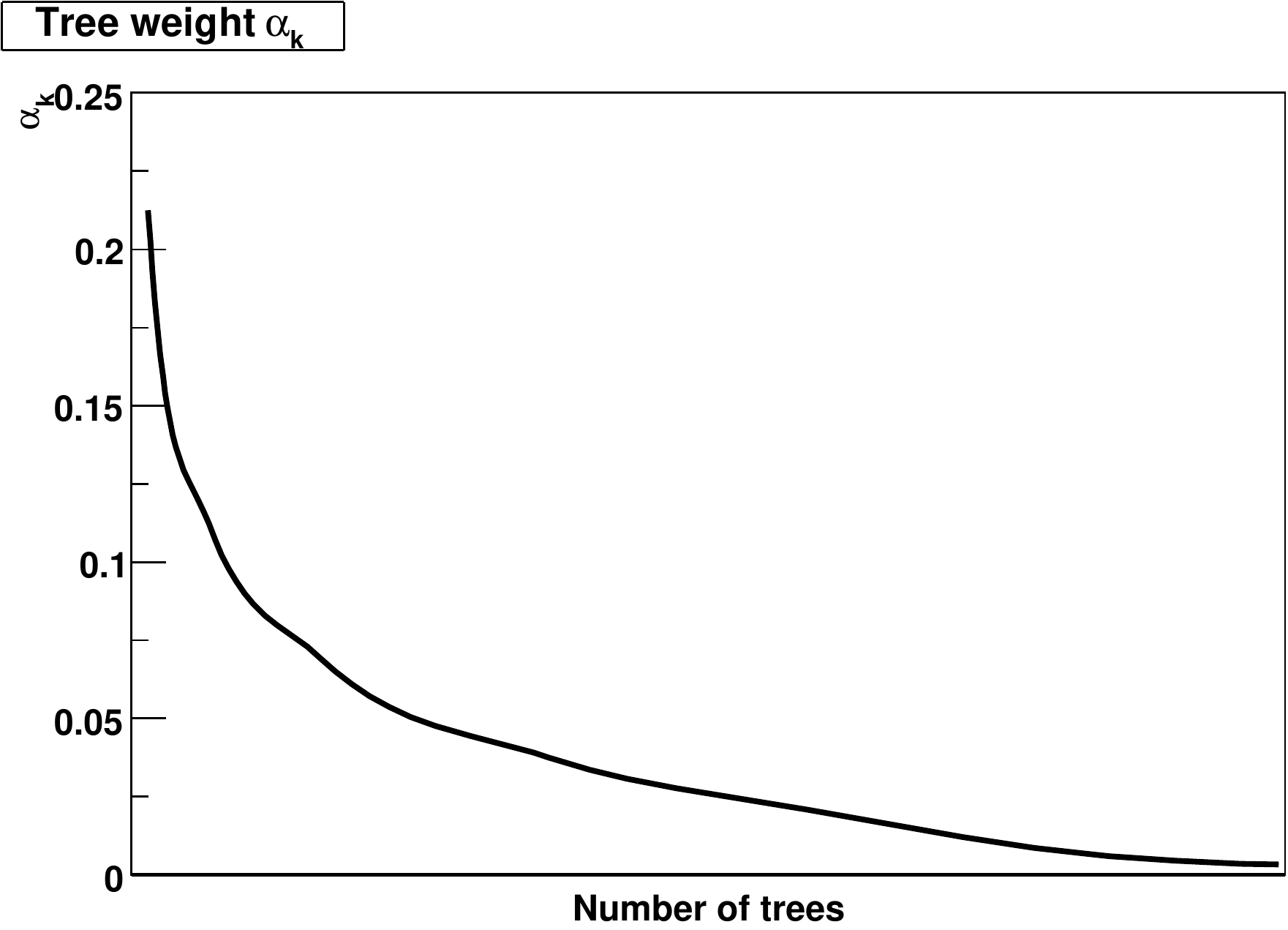}\label{fig:boost:treeweight}}
  }
  \caption{Behaviour of boosting. (a) Significance as a function of
    the number of boosted trees. (b) Signal efficiency vs. background
    efficiency for single and boosted \dts, on the training and
    testing samples. (c) Misclassification rate of each tree as a
    function of the number of boosted trees. (d) Weight of each tree
    as a function of the number of boosted trees.}
  \label{fig:boost}
\end{figure}

In a typical high-energy physics problem, the error rate may not even
be what should be optimised. A good figure of merit on the testing
sample would rather be the significance. \Fref{fig:boost:sig}
illustrates this behaviour, showing how the significance saturates
with an increasing number of boosting cycles. Arguably one
could stop before the end and save resources, but at least the
performance does not deteriorate with increasing boosting.

Another typical curve to optimise is the signal efficiency
vs. the background efficiency (the ROC curve, see \sref{sec:ROC}). \Fref{fig:boost:traintest}
clearly exemplifies this interesting property of boosted decision
trees. The performance is clearly better on the training sample than
on the testing sample (the training curves are getting very close to
the upper left corner of perfect separation), with a single tree or
with boosting, a clear sign of overtraining. But the boosted tree is
still performing better than the single tree on the testing sample,
proof that it does learn something more than memorising the training sample.

No clear explanation has emerged as to why boosting leads to such features, with
typically no loss of generalisation performance due to overtraining, but some ideas have come up. It
may have to do with the fact that during the boosting sequence, the
first tree is the best while the others are successive minor
corrections, which are given smaller weights. This is shown in
\fref{fig:boost:errorrate} and \fref{fig:boost:treeweight}, where
the misclassification rate of each new tree separately is actually
increasing, while the corresponding tree weight is decreasing. This is
not surprising: during boosting the successive trees are specialising on
specific event categories, and can therefore not perform as well on
other events. So the trees that lead to a perfect fit of the training
data are contributing very little to the final boosted \dt\ output on
the testing sample. When boosting \dts, the last tree is not an
evolution of the first one that performs better, quite the
contrary. The first tree is typically the best, while others bring
dedicated help for misclassified events. The power of boosting does
not rely in the last tree in the sequence, but rather in combining a
suite of trees that focus on different events.

A probabilistic interpretation of AdaBoost was
proposed~\cite{Proba} which gives some insight into the performance of
\bdts. It can be shown that for a boosted output $T$ flexible
enough:
\[e^{T(i)}=\frac{p(S|i)}{p(B|i)}.\]
This means that the AdaBoost algorithm will tend towards the Bayes
classifier, the maximum reachable separation.

Finally AdaBoost performance and its tendency to generalise well
despite matching very closely the training data (to the extent that in
many documented cases, to keep boosting even after the training error
has reached zero still improves the performance on the testing sample~\cite{BoostingFoundations}, in the interpolation regime~\cite{Interpolation})
have been qualitatively understood with the margins
explanation~\cite{Schapire1998,BoostingFoundations}. A classifier can
be more sure of some predictions than of others (recall
\fref{fig:limitations:distributed}), and could then generalise
better. By boosting, AdaBoost tends to increase the margins on the
training set, even after reaching zero training error.  For each
event, the margin accounts for the separability between classes,
measured by the proportion of trees that misclassify each event. For
event $x$ with truth label $y$ the margin $y\times T(x)$ for \bdt $T$
is:
\begin{align*}
  y\times T(x) & = \frac{y}{\sum_{k=1}^{N_\text{tree}}\alpha_k}\sum_{k=1}^{N_\text{tree}}\alpha_k T_k(x)\\
        & =\frac{1}{\sum_{k=1}^{N_\text{tree}}\alpha_k}\left(\sum_{k:y=T_k(x)}\alpha_k-\sum_{k:y\neq T_k(x)}\alpha_k\right),
\end{align*}
that is, the difference between the weights of single trees that
classify $x$ correctly and the weights of trees that misclassify
$x$. Boosting more means adding small corrections that tend to
increase the margin for each event. This increases the confidence in
the prediction, more likely to be correct. It makes a link with
support vector machines~\cite{Vapnik2000}, although this did not bring
great insights to improve AdaBoost in the end.

This shortcoming suggests that there may be other explanations, as
discussed in Ref.~\cite{Interpolation}, focusing on the interpolation
regime when the training error has already reached zero but boosting
further still leads to testing error improvement (better
generalisation). The combination of large trees focusing on extremely
local neighbourhoods of the training dataset and averaging over a
large number of trees seems to prevent overfitting efficiently. This
has been interpreted in the more general framework of double descent
risk curve~\cite{Belkin2019}. With boosting, the interpolating regime
behaviour (see \fref{fig:overtraining:interpolation}) may kick in even
before the interpolating threshold, possibly explaining why typical
\bdt training curves look like \fref{fig:overtraining:flat}.

\subsection{Gradient boosting}
\label{sec:gradboost}
While trying to understand how AdaBoost and
other boosting algorithms work, they were originally recast in the
statistical framework of arcing algorithms (an acronym for adaptive
reweighting and combining)~\cite{Breiman1998,Breiman1999}. At each
step, a weighted minimisation is performed followed by a recomputation
of the classifier and weighted input. This was further developed to
become gradient boosting~\cite{friedman2001}. Boosting is formulated as a
numerical optimisation problem, trying to minimise the loss function
by adding trees using a gradient descent procedure rather than giving
a higher weight to misclassified events.

Formally, consider a model $F$ built iteratively, its imperfect
instance at step $k$ being $F_k$. $F_k$ is therefore an approximation of the best possible model (in some cases $F_k(x)\neq y$), which is to be improved at the next iteration. This is achieved by adding a new component $h_k$ such that:
\[F_{k+1}(x) = F_k(x) + h_k(x) = y,\]
or equivalently:
\[ h_k(x) = y - F_k(x).\]
Rather than training $F_{k+1}$ a new classifier can be trained to fit the residual $y - F_k(x)$, which
corresponds to the part that the current model $F_k$ cannot treat
correctly. If $F_{k+1}(x)$ is still not satisfactory, new iterations
can be fitted.

The link with gradient descent is explicit when considering the
particular case of the mean squared error (MSE) loss function (a
typical case for regression problems, see
\sref{sec:regression}):
\[L_\text{MSE}(x,y) = \frac{1}{2}\left(y-F_k(x)\right)^2.\] Minimising
the loss $J=\sum_iL_\text{MSE}(x_i,y_i)$ by adjusting all $F_k(x_i)$
leads to:
\[\frac{\partial J}{\partial F_k(x_i)} = \frac{\partial L_\text{MSE}(x_i,y_i)}{\partial F_k(x_i)} = F_k(x_i) - y_i.\]
Residuals can therefore be interpreted as negative gradients:
\[h_k(x_i) = y_i - F_k(x_i) = -\frac{\partial J}{\partial F_k(x_i)}.\]
The concept can be generalised to any differentiable loss function
instead of MSE. For instance AdaBoost corresponds to an exponential
loss $e^{-F_k(x)y}$.

There are several variants of gradient boosting algorithms on the
market. Techniques presented in \sref{sec:Others} with
subsampling of the training set and tree parameters can be used (in
particular a bagging-like approach without replacement), leading to
stochastic gradient boosting~\cite{Friedman2002}. These regularisation
techniques help prevent overfitting.

\subsection{Boosting examples}
\label{sec:examples}
The examples of this section illustrate typical behaviours of \bdts.

\subsubsection{The XOR problem}
\begin{figure}
  \centerline{
    \subfigure[]
    {\includegraphics[width=.45\textwidth]{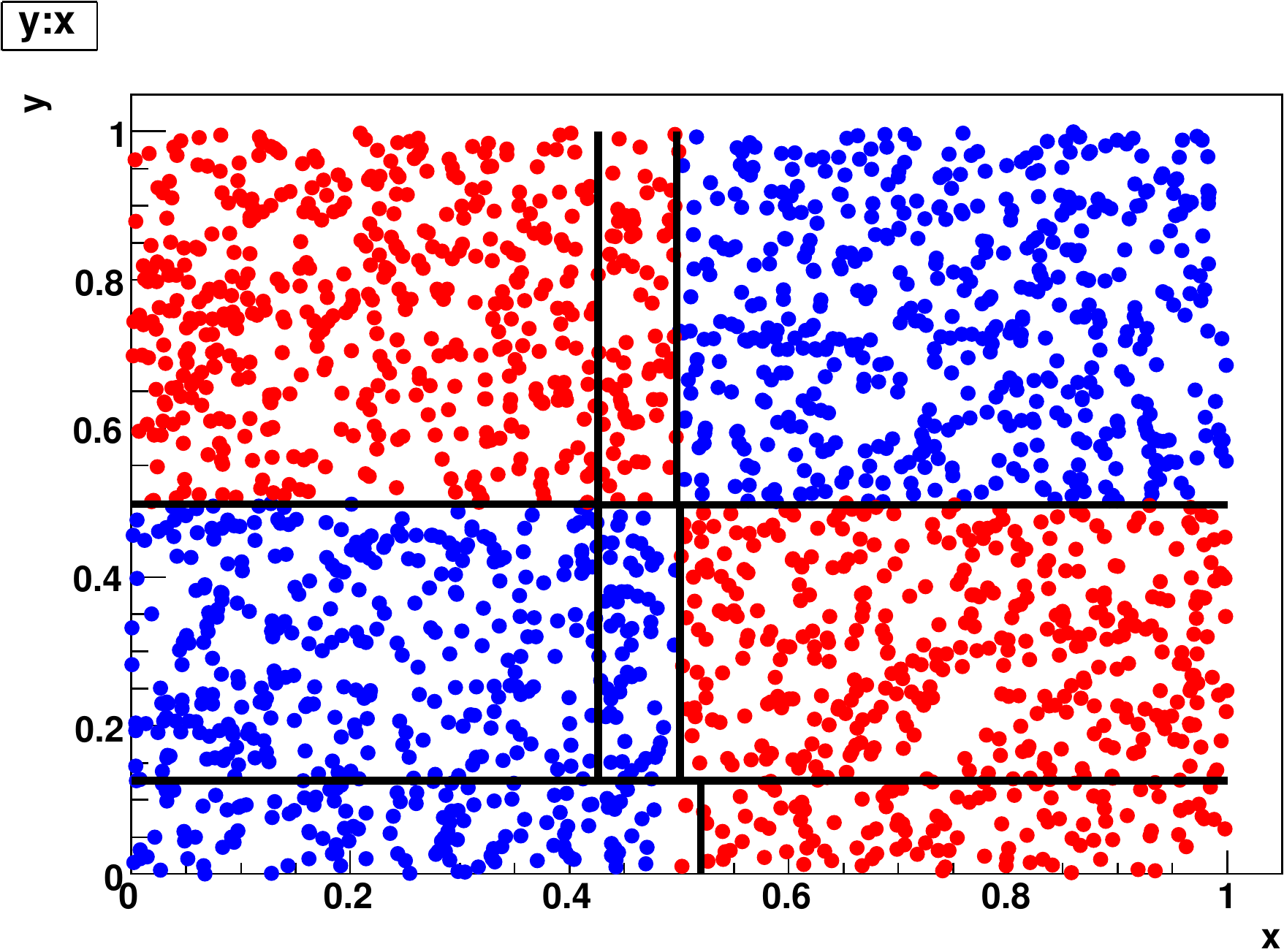}\label{fig:xor:cuts}}
    \hspace*{20pt}
    \subfigure[]
    {\includegraphics[width=.45\textwidth]{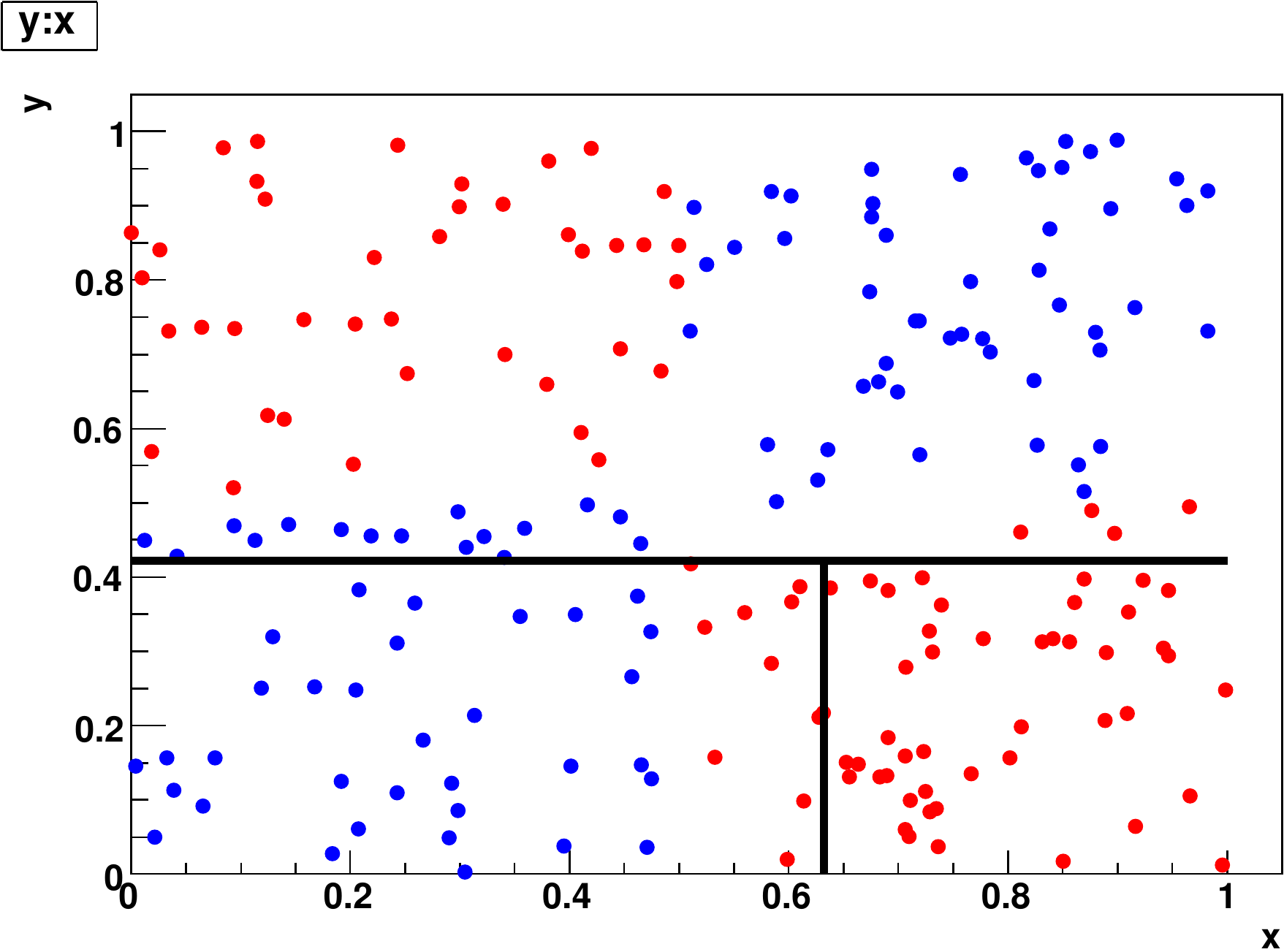}\label{fig:xor:cuts100}}
  }
  \centerline{
    \subfigure[]
    {\includegraphics[width=.45\textwidth]{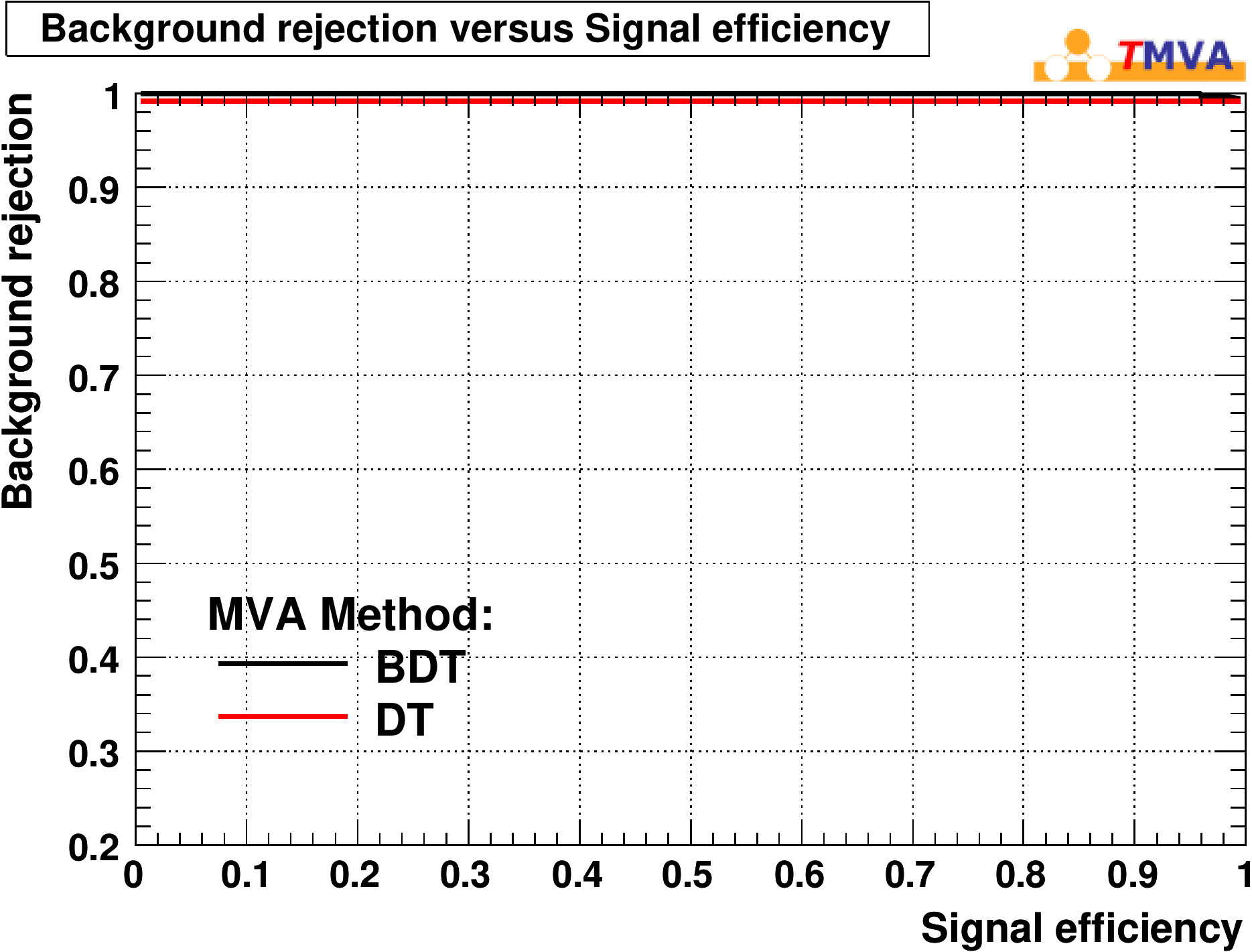}\label{fig:xor:roc}}
    \hspace*{20pt}
    \subfigure[]
    {\includegraphics[width=.45\textwidth]{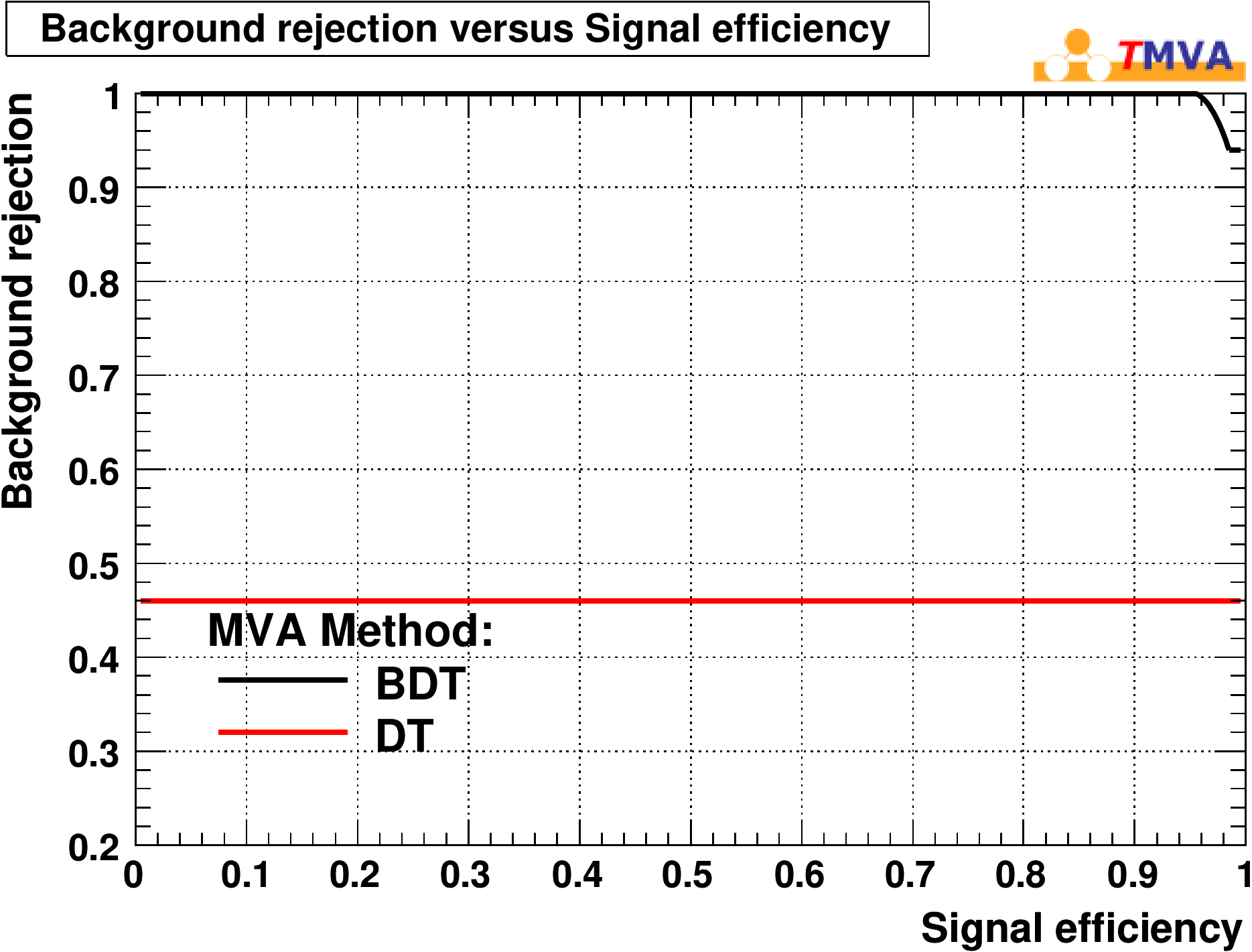}\label{fig:xor:roc100}}
  }
  \caption{The XOR problem. Signal is in blue, background in red. The
    left column (a and c) uses sufficient statistics, while the right
    column has a limited number of training events. The top plots (a
    and b) show the signal and background distributions as well as the
    criteria used by the first \dt. Bottom plots (c, d) illustrate the
    background rejection vs. signal efficiency curves for the first
    \dt\ (red) and for the boosted \dts\ (black), all run on the same testing
    events.}
  \label{fig:xor}
\end{figure}

The XOR problem is a small version of the checkerboard, illustrated in
\fref{fig:xor}. With enough statistics (\fref{fig:xor:cuts}
and \fref{fig:xor:roc}), even a single tree is already able to find
more or less the optimal separation, so boosting cannot actually do
much better.

The exercise can be repeated, this time with limited statistics
(\fref{fig:xor:cuts100} and \fref{fig:xor:roc100}). Now a single tree
is not doing such a good job anymore. Boosted \dts, on the other hand, are doing
almost as well as with full statistics, separating almost perfectly
signal and background. This illustrates very clearly how the
combination of weak classifiers (see for instance the lousy
performance of the first tree) can generate a high performance
discriminant with a boosting algorithm.

\subsubsection{Number of trees and overtraining}
\label{sec:ntrees}
\begin{figure}
  \centerline{
    \subfigure[]
    {\includegraphics[height=.38\textwidth]{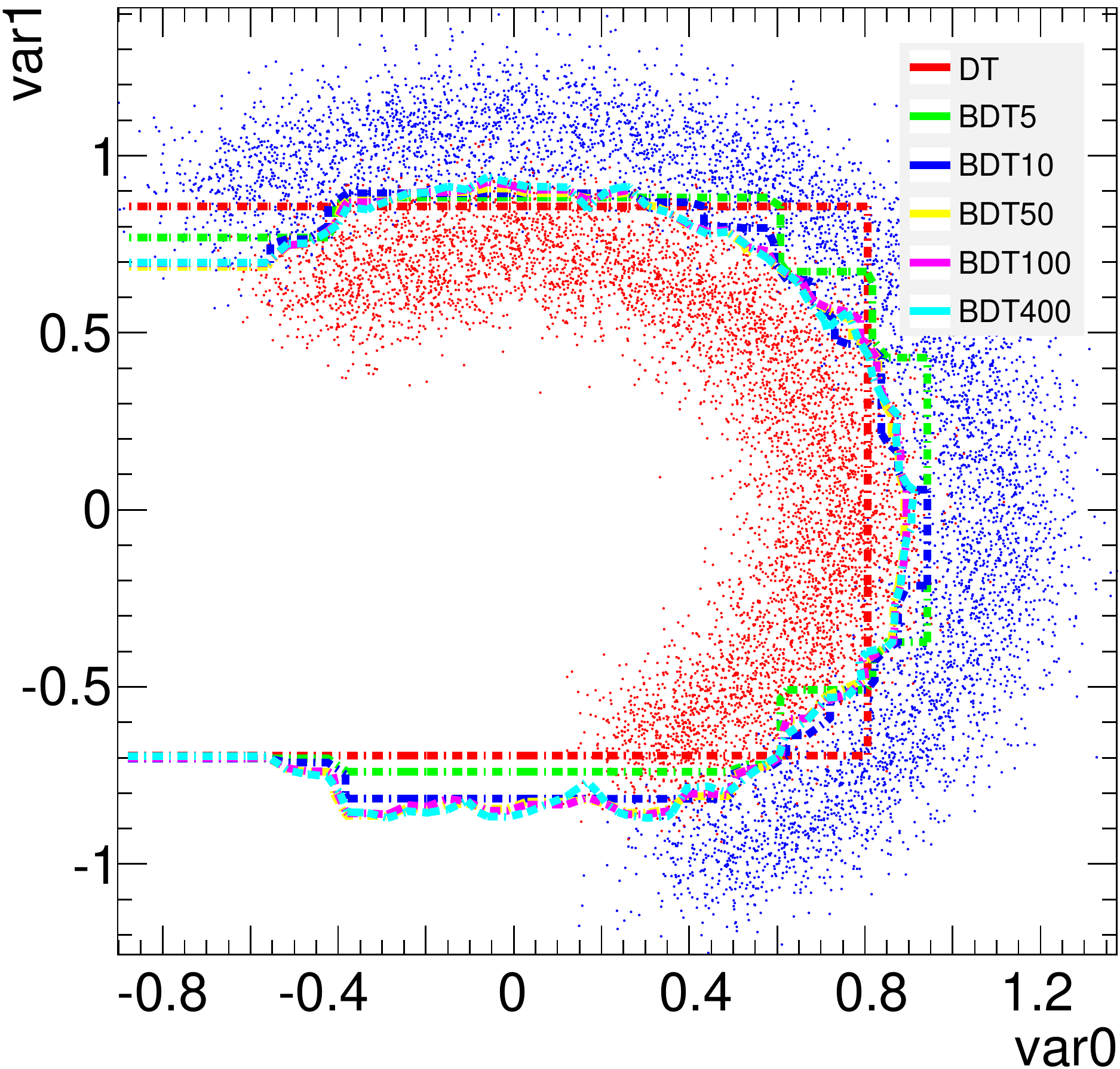}\label{fig:circ:2D}}
    \hspace*{20pt}
    \subfigure[]
    {\includegraphics[height=.38\textwidth]{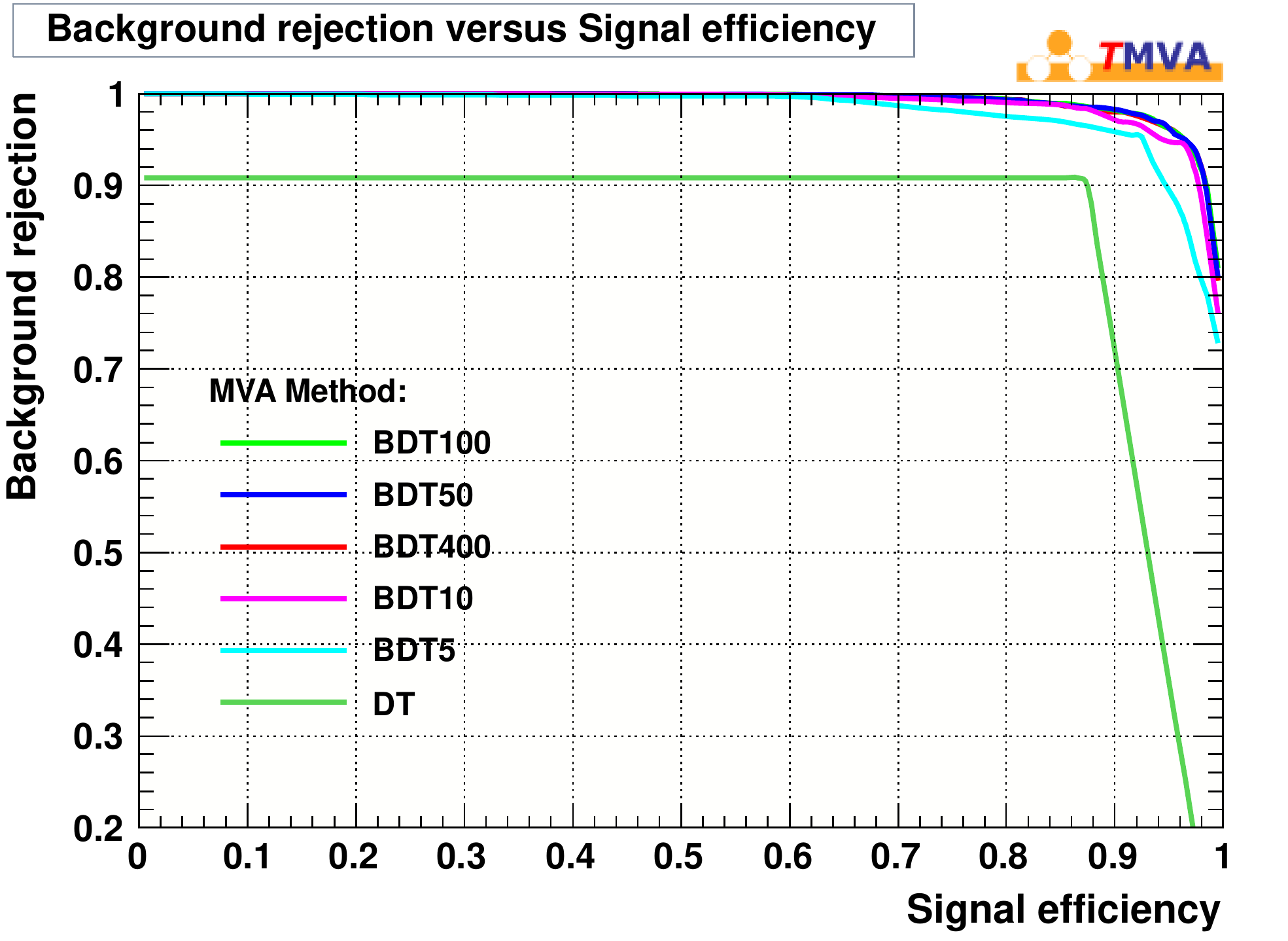}\label{fig:circ:roc}}
  }
  \caption{(a) 2D dataset and decision contour corresponding to
    several discriminants. (b) Background rejection vs. signal
    efficiency curves for a single \dt
    (dark green) and \bdts\ with an increasing number of trees (5 to
    400).}
  \label{fig:circ}
\end{figure}

This example uses a highly correlated dataset, shown in \fref{fig:circ:2D}.

Figure \ref{fig:circ:roc} compares the performance of a
single \dt\ and boosted \dts\ with an
increasing number of trees (from 5 to 400). All other parameters
are kept to their default value in the TMVA package~\cite{TMVA}.
The performance of the single tree is not so good, as
expected since the default parameters make it very small, with a
depth of 3 (it should be noted that a single bigger tree could solve
this problem easily). Increasing the number of trees improves the
performance until it saturates in the high background rejection and
high signal efficiency corner. Adding more trees does not seem to
degrade the performance, the curve stays in the optimal
corner. Looking at the contours in \fref{fig:circ:2D} it wiggles a little for larger \bdts, as they tend to
pick up features of the training sample. This is overtraining.

\begin{figure}
  \centerline{
    \subfigure[Single \dt]
    {\includegraphics[width=.31\textwidth]{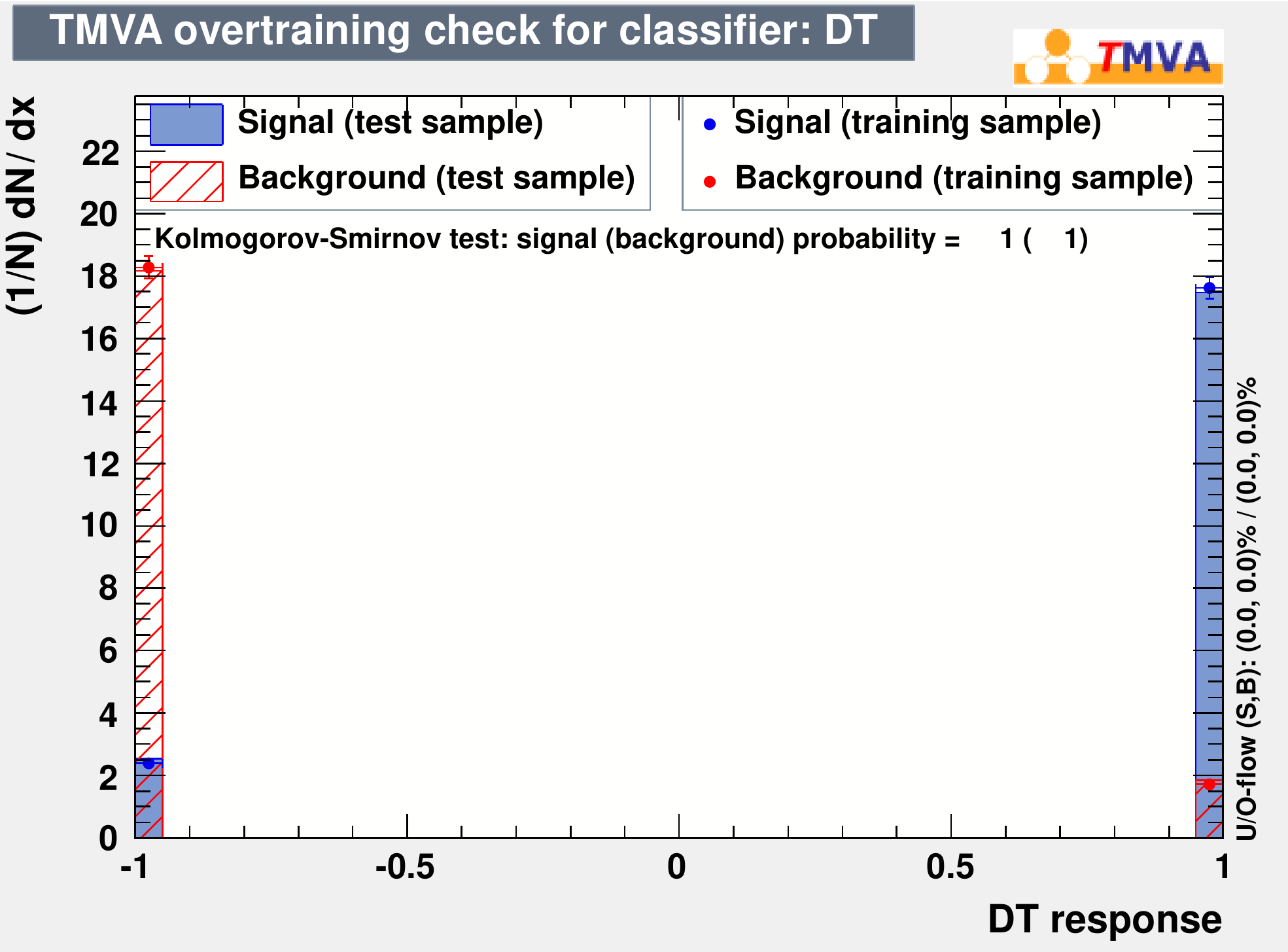}\label{fig:circoutput:DT}}
    \hspace*{4pt}
    \subfigure[5 trees]
    {\includegraphics[width=.31\textwidth]{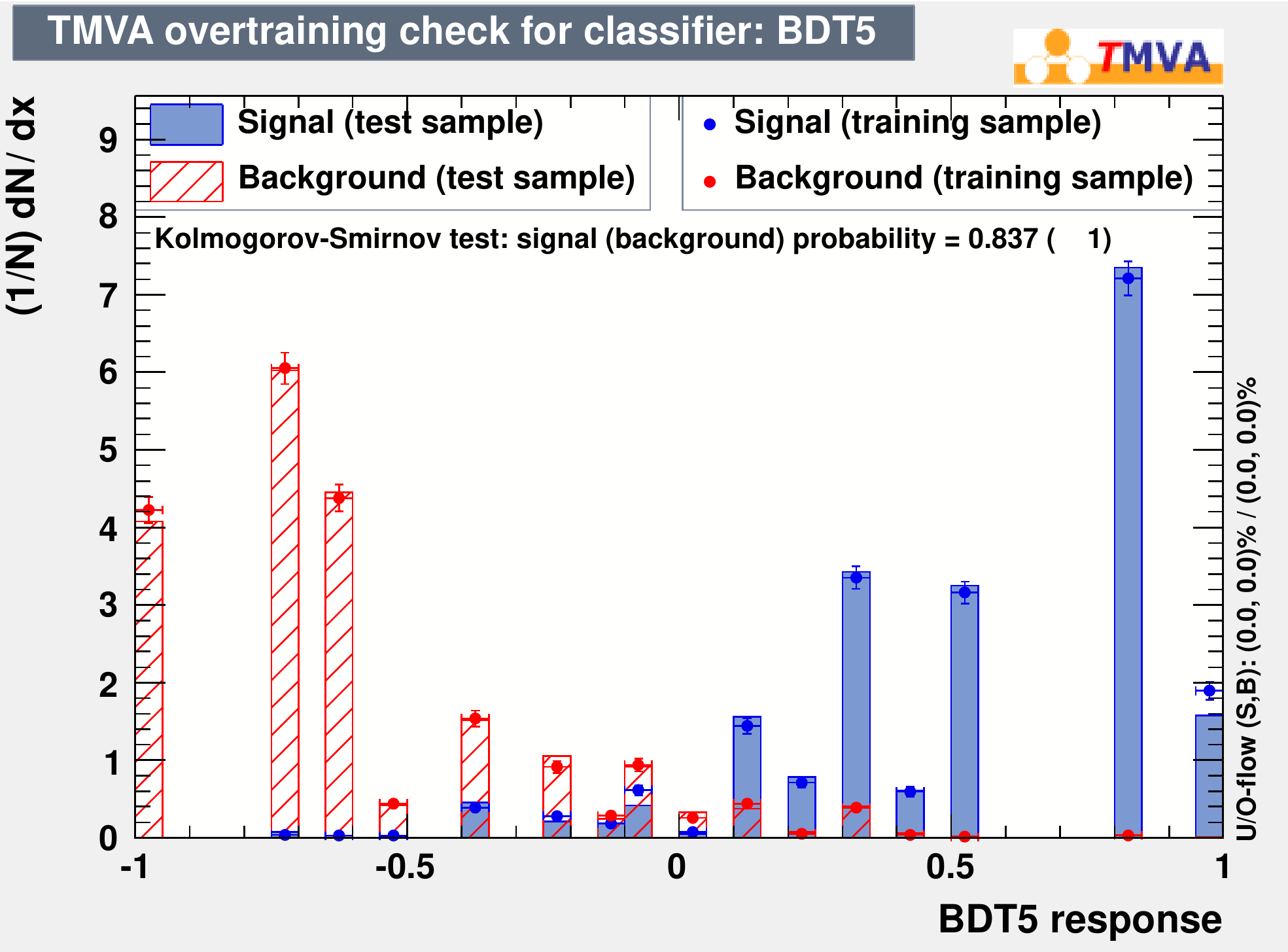}\label{fig:circoutput:BDT5}}
    \hspace*{4pt}
    \subfigure[10 trees]
    {\includegraphics[width=.31\textwidth]{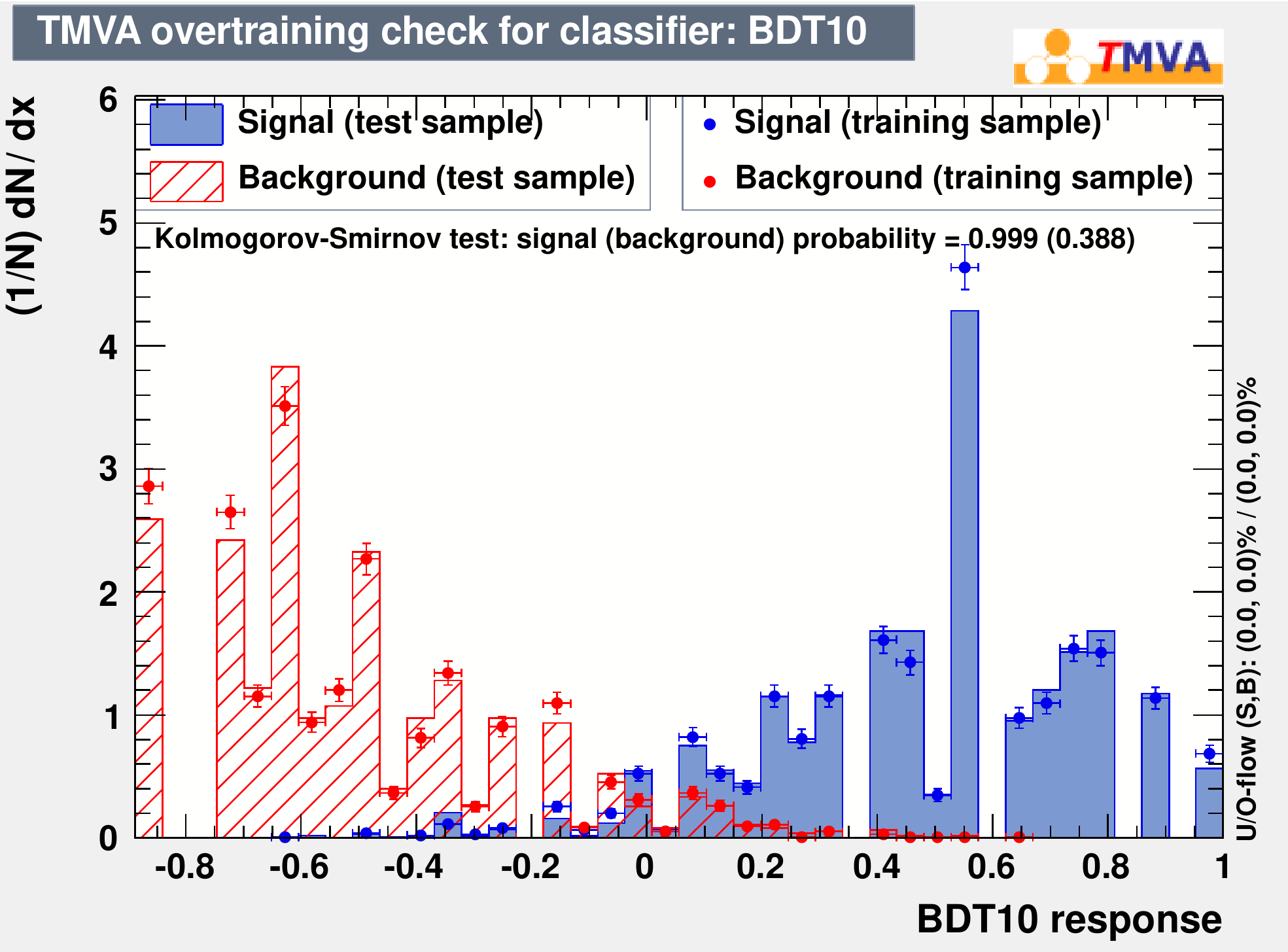}\label{fig:circoutput:BDT10}}
  }
  \centerline{
    \subfigure[50 trees]
    {\includegraphics[width=.31\textwidth]{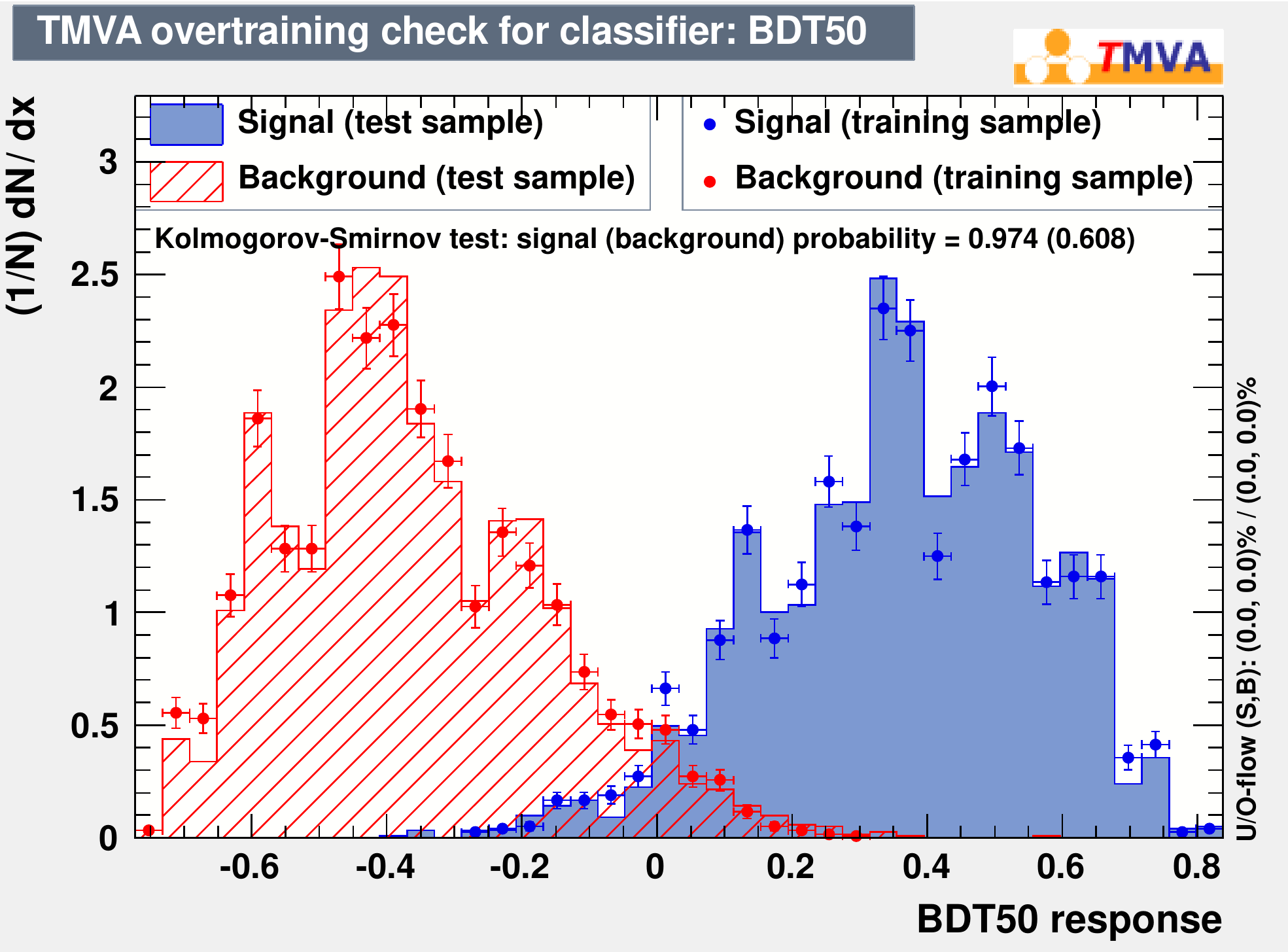}\label{fig:circoutput:BDT50}}
    \hspace*{4pt}
    \subfigure[100 trees]
    {\includegraphics[width=.31\textwidth]{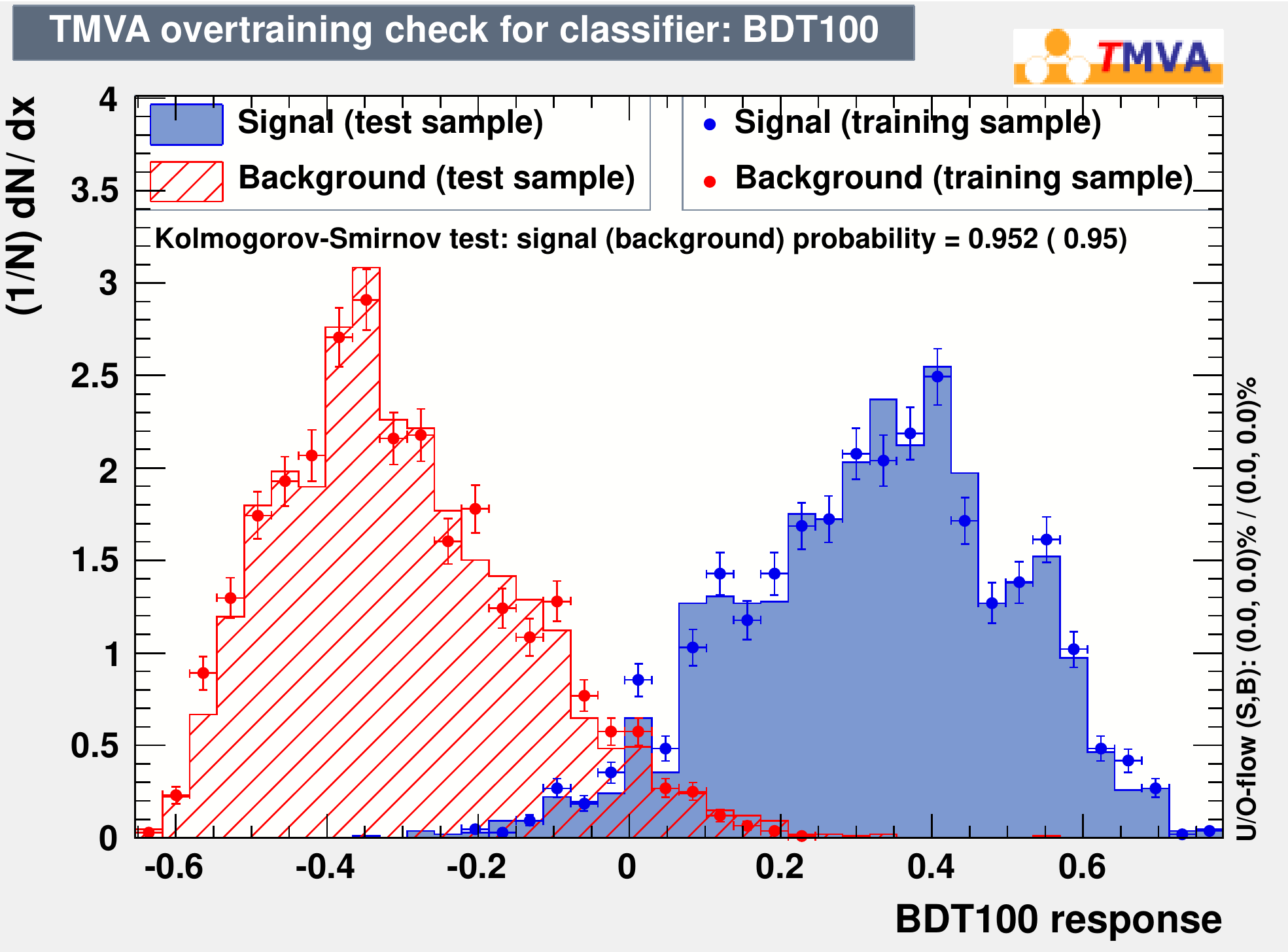}\label{fig:circoutput:BDT100}}
    \hspace*{4pt}
    \subfigure[400 trees]
    {\includegraphics[width=.31\textwidth]{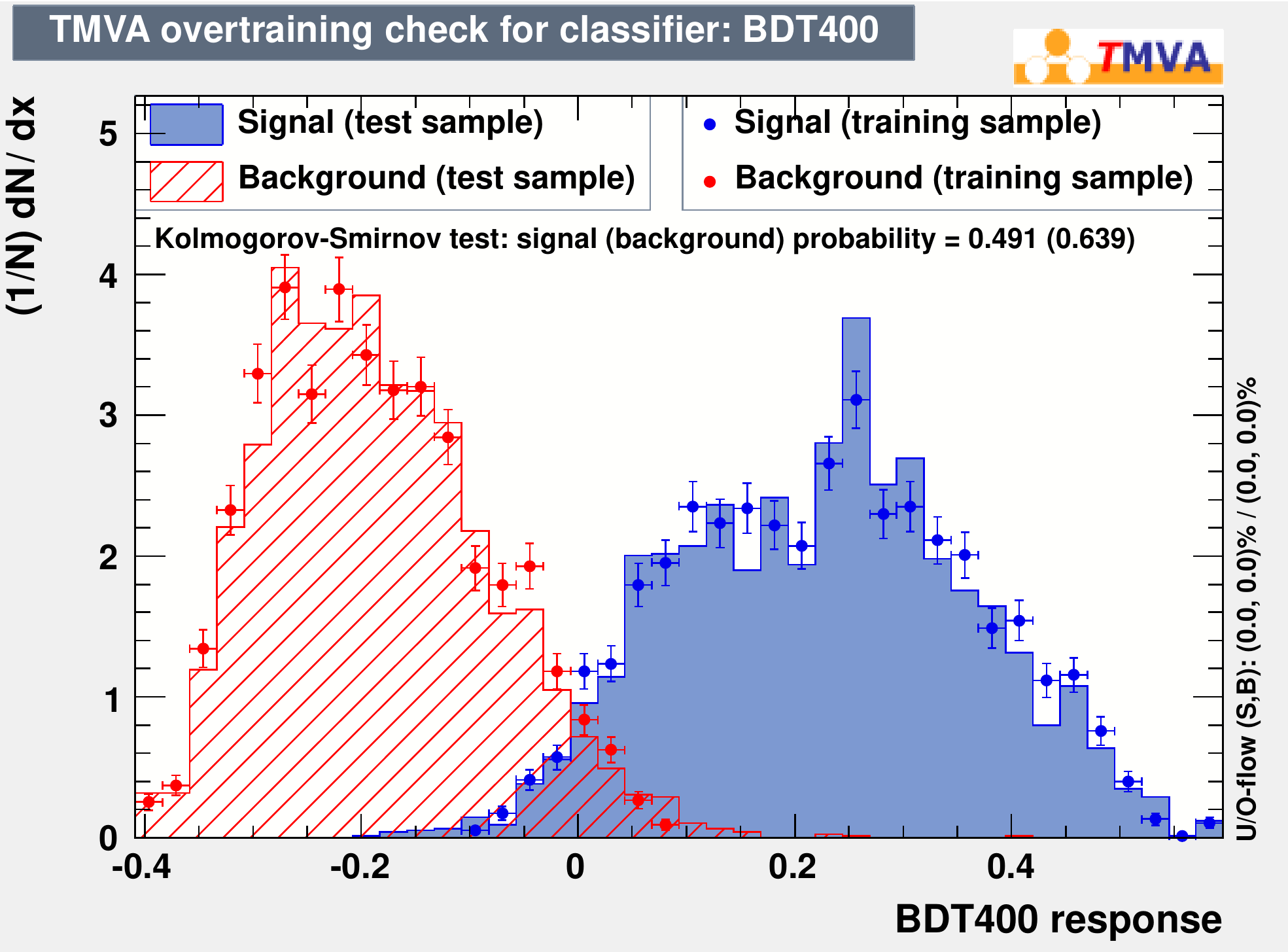}\label{fig:circoutput:BDT400}}
  }
  \caption{Comparison of the output on training (markers) and testing
    (histograms) signal (blue) and background (red) samples for \bdts\ with 1, 5, 10, 50, 100 and 400
    trees (from top left to bottom right). The Kolmogorov--Smirnov test quantifies the (dis)agreement between training and testing outputs.}
  \label{fig:circoutput}
\end{figure}

Another sign of overtraining also appears in \fref{fig:circoutput},
showing the output of the various \bdts\ for signal and background,
both on the training and testing samples: larger \bdts\ tend to show
differences between the two samples (as quantified by a
Kolmogorov--Smirnov (KS) test in the figures, especially
\fref{fig:circoutput:BDT400}), as they adjust to peculiarities of the
training sample that are not found in an independent testing
sample. The output acquires a `better' shape with
more trees, really becoming quasi-continuous, which would allow to cut
at a precise efficiency or rejection.

\begin{figure}
  \centerline{
    \subfigure[]
    {\includegraphics[width=.31\textwidth]{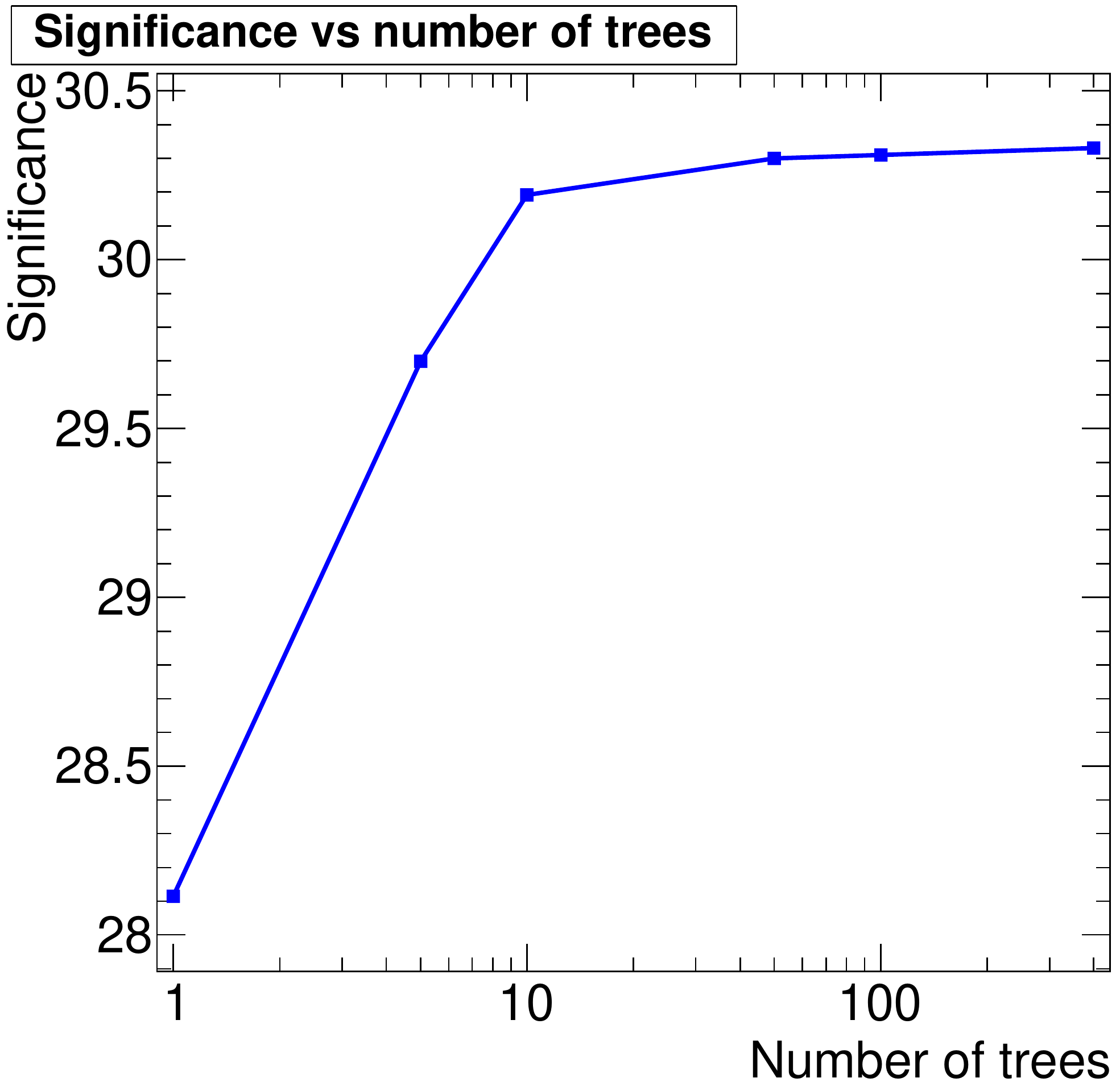}\label{fig:circtreeweight:sig}}
    \hspace*{4pt}
    \subfigure[]
    {\includegraphics[width=.31\textwidth]{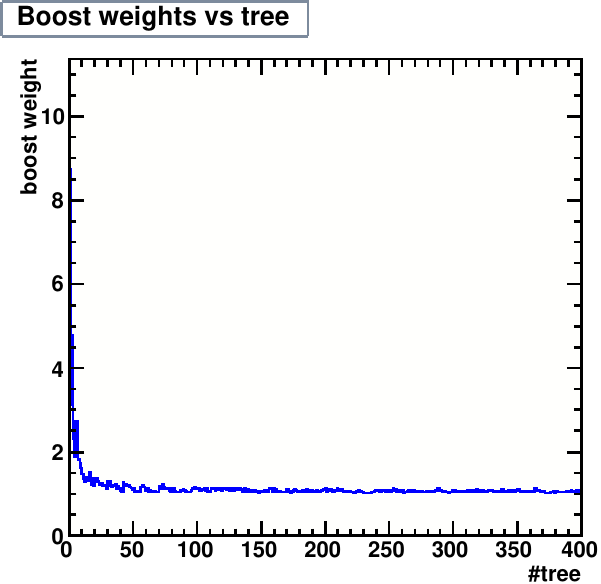}\label{fig:circtreeweight:boostweight}}
    \hspace*{4pt}
    \subfigure[]
    {\includegraphics[width=.31\textwidth]{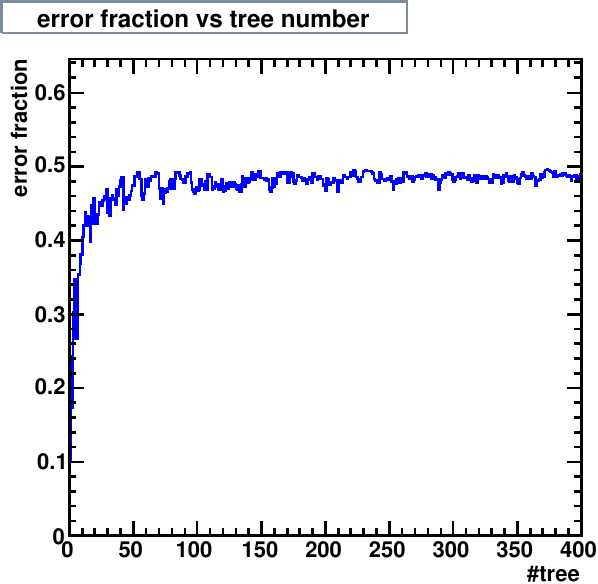}\label{fig:circtreeweight:treeweight}}
  }
  \caption{(a) Maximum significance of all \bdts. (b) Boost weight of
    each tree. (c) Error fraction of each tree (0.5 means random
    guessing).}
  \label{fig:circtreeweight}
\end{figure}

Both figures do exhibit clear signs of overtraining, but is it really
an issue? As mentioned before (see \sref{sec:overtraining}) what
really matters in the end is the performance in data analysis and on
the testing sample. One way to evaluate this is to compute the maximum
significance $s/\sqrt{s+b}$ (see \sref{sec:significance}). It is shown
in \fref{fig:circtreeweight:sig} for the same \bdts\ as shown in
\fref{fig:circoutput}, with increasing number of trees. The best
significance is actually obtained with the 400-tree \bdt, following
what was described at the end of \sref{sec:adaboost}. To be fair, the
performance is very similar already with 10 trees. Now, comparing the
outputs in \fref{fig:circoutput}, if interested in a smoother result,
10 trees might not be enough, but 50 would probably do, without the
overhead of eight times more trees. Such a choice should in any case
not be made based on overtraining statements comparing performance on
the training and testing samples (as some are tempted to do, seeing an
increasing disagreement, quantified by the KS test, between outputs on the training and testing
samples), but rather on final expected physics performance (the final number of the analysis, for instance the significance from the complete statistical analysis, possibly including systematic uncertainties). \BDTs are often in
the situation described in \fref{fig:overtraining:flat}, meaning that
their performance is not decreasing when boosting longer, even as the
discrepancy in performance between train and test keeps increasing.

This example also illustrates the performance of each tree in a
boosting sequence. \Fref{fig:circtreeweight:boostweight} shows
the rapid decrease of the weight $\alpha_k$ of each tree, while at the
same time the corresponding misclassification rate $\varepsilon_k$ of
each individual tree increases rapidly towards just below 50\%, that
is, random guessing (\fref{fig:circtreeweight:treeweight}). It
confirms that the best trees are the first ones, while the others are
only minor corrections.

\subsection{Other boosting algorithms}
\label{sec:otherboost}
AdaBoost is but one of many boosting algorithms. It is also referred
to as discrete AdaBoost to distinguish it from other AdaBoost
flavours. The Real AdaBoost algorithm~\cite{Proba} defines each \dt\
output as:
\[T_k(i)=0.5\times\ln\frac{p_k(i)}{1-p_k(i)},\] 
where $p_k(i)$ is the purity of the leaf on which event $i$ falls.
Events are reweighted as:
\[ w_i^k \to w_i^{k+1}=w_i^k\times e^{-y_i T_k(i)}\]
and the boosted result is $T(i) = \sum_{k=1}^{N_\text{tree}}T_k(i)$.
Gentle AdaBoost and LogitBoost (with a logistic function)~\cite{Proba}
are other variations.

$\varepsilon$-Boost, also called shrinkage~\cite{friedman2001}, consists in
reweighting misclassified events by a fixed factor $e^{2\varepsilon}$
rather than the tree-dependent $\alpha_k$ factor of AdaBoost.
$\varepsilon$-LogitBoost~\cite{Proba} is reweighting them with a
logistic function $\frac{e^{-y_iT_k(i)}}{1+e^{-y_iT_k(i)}}$.
$\varepsilon$-HingeBoost~\cite{Miniboone2} is only dealing with
misclassified events:
\[ w_i^k \to w_i^{k+1}=\mathbb{I}(y_i\times T_k(i)\le 0).\]

Finally the adaptive version of the `boost by
majority'~\cite{Freund} algorithm is called BrownBoost~\cite{Brown}. It
works in the limit where each boosting iteration makes an
infinitesimally small contribution to the total result, modelling this
limit with the differential equations that govern Brownian motion.

\subsection{Boosted regression trees}
\label{sec:regression}
From their very introduction~\cite{Breiman}, trees have been
considered for classification (\dts) and for regression (regression
trees), where instead of identifying `signal-like' or
`background-like' regions of phase space, tree leaves each contain a
single real value supposed to approach the target function.

During tree building for regression, the maximisation of the decrease
of impurity in \dts is replaced by the reduction of the standard
deviation or of the mean squared error:
\[d(t) = \frac{1}{N_t} \sum_{N_t}(y-\hat y_t)^2,\]

for a node $t$ with $N_t$ events, regression target $y$ of each event
in the node and mean value $\hat y_t$ of regression targets of all
events in the node. Another typical choice for $d$ is the mean
absolute error:
\[\frac{1}{N_t} \sum_{N_t}\left|y- \text{median}(y)_t\right|.\]

Constructing a regression tree is about finding the attribute that
return the highest reduction in $d$ (i.e., the most homogeneous nodes)
when going from node $t$ to nodes $t_P$ and $t_F$ (see
\sref{sec:split} for notations):
\[\Delta d(S,t)=d(t)- p_P\cdot d(t_P) - p_F\cdot d(t_F).\]

The splitting stops when nodes become too small or when their internal
variation is sufficiently small. The regression tree output is the
mean (or median if using the mean absolute error) value of the
training events in the corresponding (leaf) node. So a regression tree
partitions the feature space of input variables into hyperrectangles
and then fits a constant inside each box.

When boosting regression trees, there are no longer properly and
wrongly classified events, so the misclassification rate cannot be
computed to reweight events. Instead the average loss
$\langle L^k\rangle$ after the $k^\text{th}$ tree is computed over the
training sample, and the boosting quantity
$\beta_k=\langle L^k\rangle/\left(1-\langle L^k\rangle\right)$ is
derived. The reweighting of events is then computed based on their
individual loss $L^k(i)$:
\[ w_i^k \to w_i^{k+1}=w_i^k\times \beta_k^{1-L^k(i)}. \]

The training process is then similar to that of \bdts, and the final
prediction of the fitted value is the weighted average of all tree
outputs.

\subsection{\BDTs\ in high-energy physics}
\label{sec:hep}
\BDTs have become very popular in high-energy physics. A few usage
examples are presented in \sref{sec:UseCases}. Their proper usage also
means addressing issues linked to systematic uncertainties, as
reported in \sref{sec:syst}.

\subsubsection{Use cases}
\label{sec:UseCases}
The MiniBooNe experiment at Fermilab, searching for neutrino
oscillations, was the first in the field to compare the performance of
different boosting algorithms and artificial neural networks for
analysis and particle identification~\cite{Miniboone1,Miniboone2}, on
Monte Carlo samples. Trees with up to 120 variables were tested, with
different boosting algorithms and up to thousands of trees. These
studies introduced \bdts in the particle physics world.

\begin{figure}
  \centerline{
    \subfigure[]
    {\includegraphics[width=.49\textwidth]{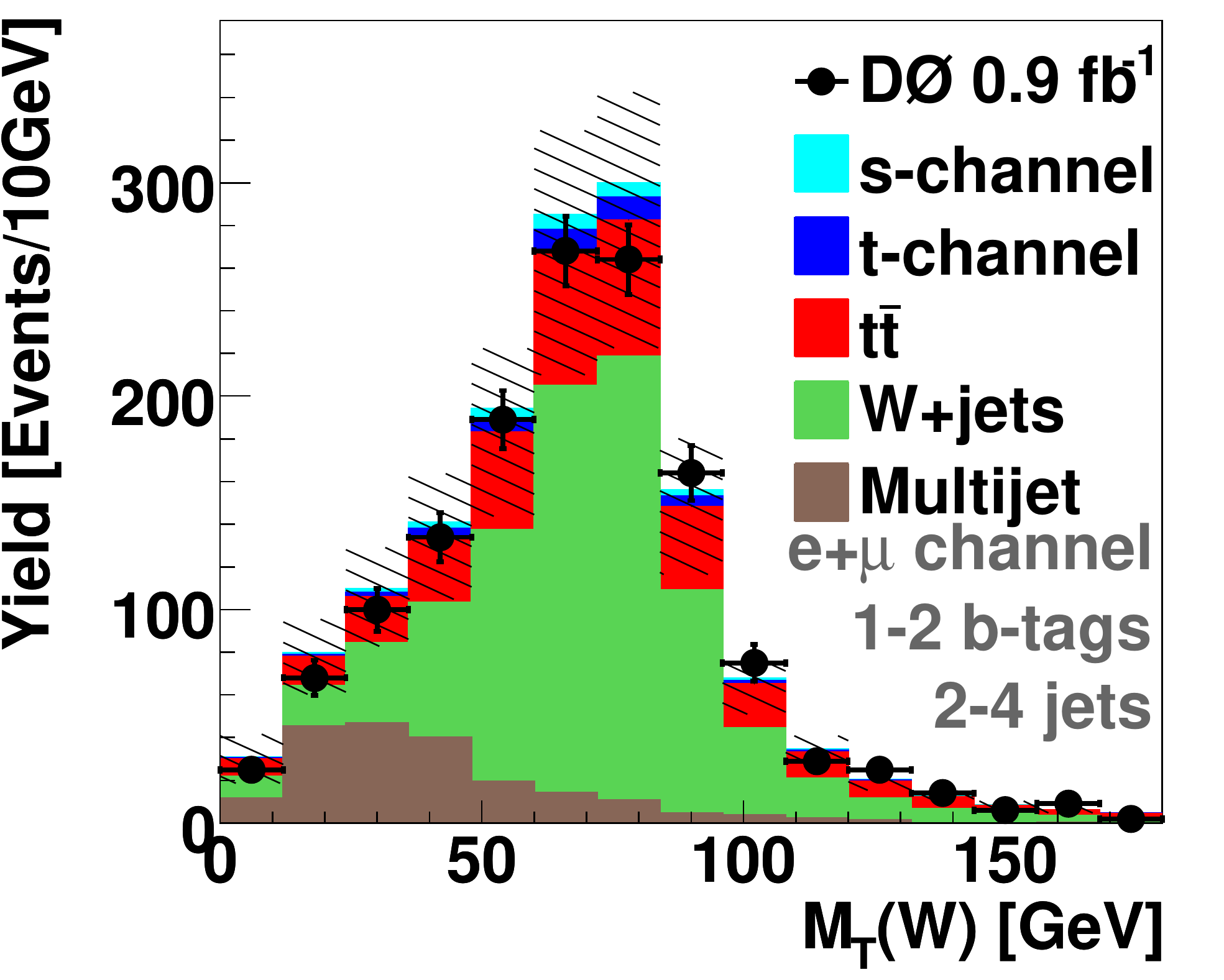}\label{fig:singletop:mtw}}
    \hspace*{10pt}
    \subfigure[]
    {\includegraphics[width=.49\textwidth]{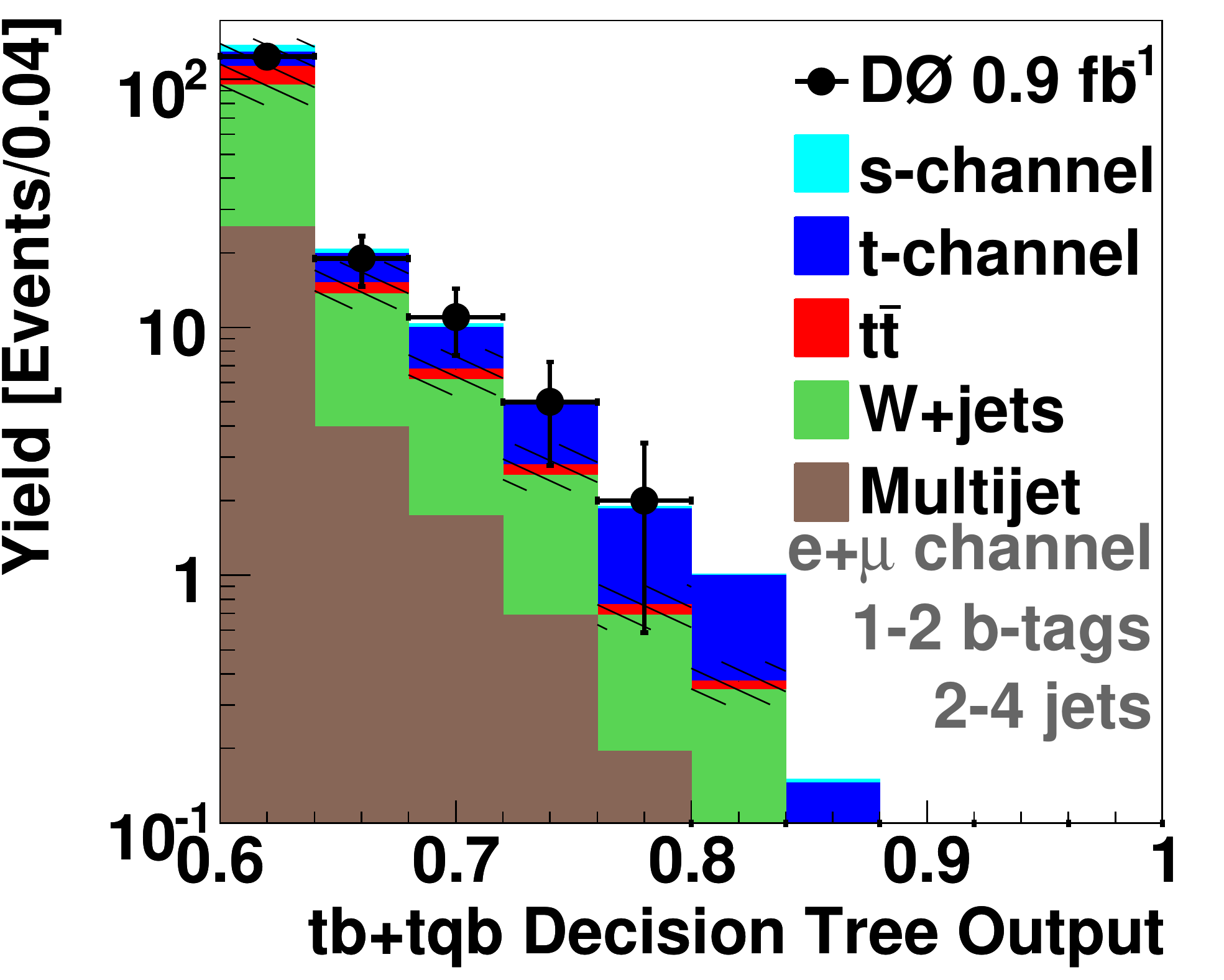}\label{fig:singletop:bdt}}
  }
  \centerline{
    \subfigure[]
    {\includegraphics[width=.49\textwidth]{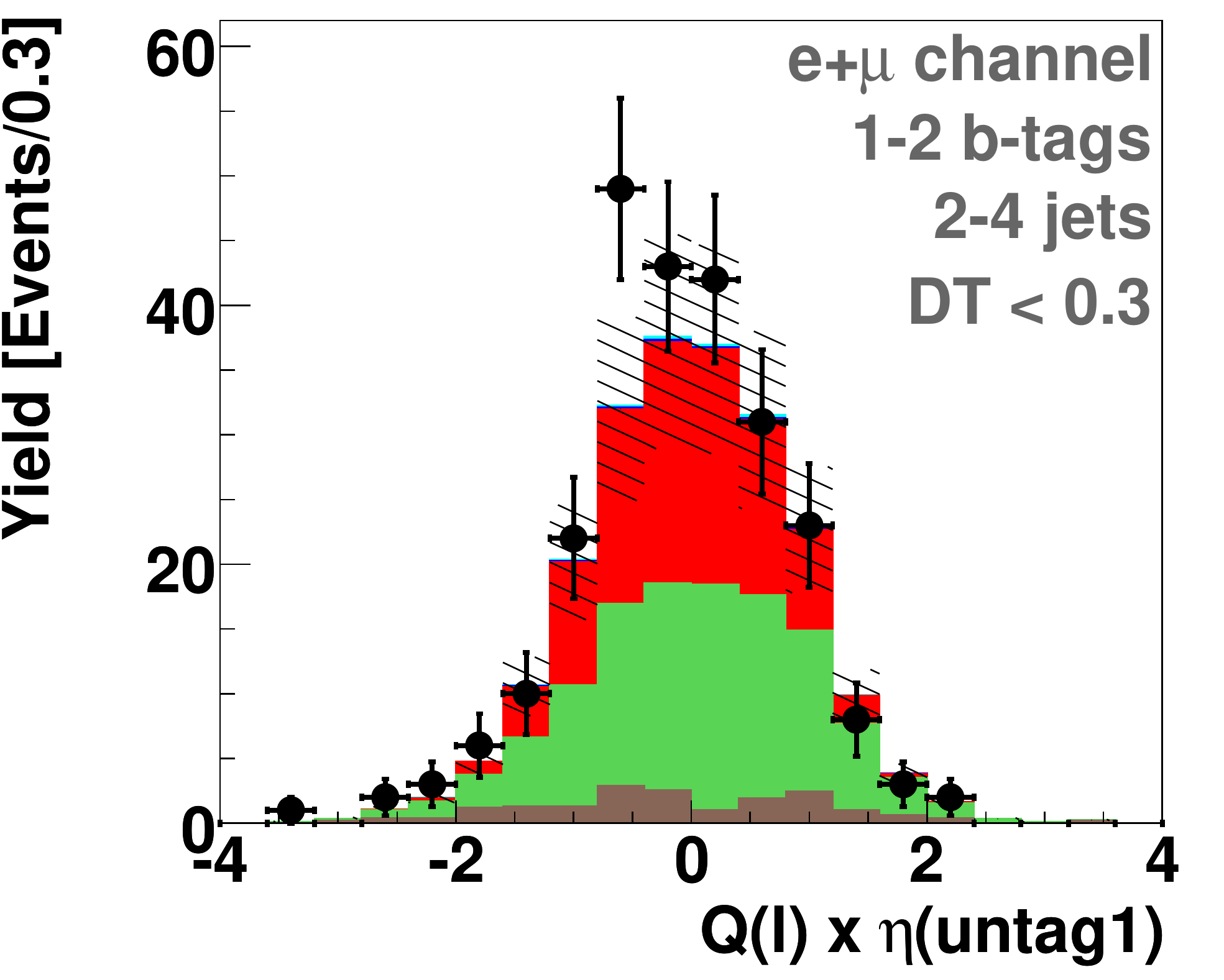}\label{fig:singletop:qetalow}}
    \hspace*{10pt}
    \subfigure[]
    {\includegraphics[width=.49\textwidth]{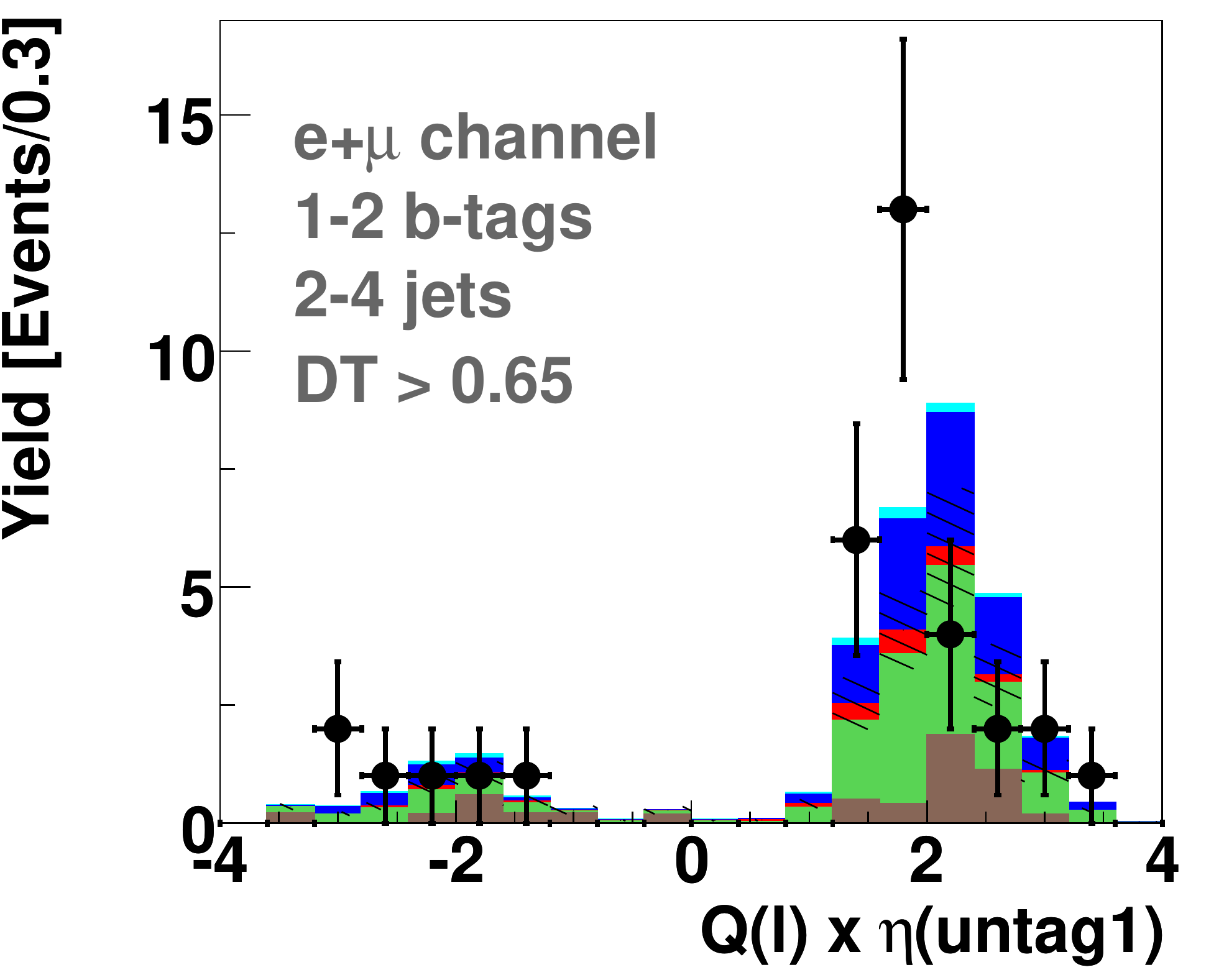}\label{fig:singletop:qetahigh}}
  }
  \caption{Usage of \bdts in physics analysis~\cite{D02}. (a) A discriminating variable, with uncertainty larger than the expected signal (in blue). (b) The \bdt output with much smaller uncertainty. (c, d) Discriminating variable when selecting only events with low (high) \bdt output, showing background (signal)-like shape.}
  \label{fig:singletop}
\end{figure}

The concept of \bdts was picked up by the D0 experiment at Fermilab,
leading to the first evidence (and then observation) of single top
quark production in Tevatron data~\cite{D01,D02}. Among the 49
variables used, some had very similar definitions (like the scalar sum
of transverse momentum of various jets), which was beneficial as not
all of them suffer from the same mismeasurements on an event-by-event
basis. \BDTs happened to perform slightly better than two other
techniques used: the matrix element calculation and Bayesian neural
networks. Without such advanced techniques, the signal could not have
been seen with the dataset available at the time: the total
uncertainty on the model prediction was much larger than the expected
signal, as illustrated in \fref{fig:singletop:mtw}. This also means
that no single distribution (apart from the \bdt output shown in
\fref{fig:singletop:bdt}) could really show the new observed process,
leading to scepticism in the community (`I want to see a mass peak!'
is a common argument, reflecting on the fact that people are more
confident in the result if they can see the signal in a physical
distribution). Various cross-checks were performed to increase the
degree of belief in the final outcome (removing top-quark-mass-related variables
during training, validating the description of the \bdt output in
regions depleted in signal, analysing the shape of other variables
after selecting low or high \bdt output events enriched in background
or signal events as shown in \fref{fig:singletop:qetalow} and
\fref{fig:singletop:qetahigh}, respectively, etc.).

Since then, \bdts have become a bread and butter technique in high
energy physics and are extensively used in physics analyses (to extract
their tiny signal from large backgrounds or distinguish between
different signals) and object
identification at the Tevatron or the LHC. In the ATLAS experiment
$\tau$-lepton identification~\cite{ATLAS-CONF-2017-029} and
flavour tagging~\cite{FTAG-2018-01} used \bdts in Run 2, and the $\tau$-lepton
energy is estimated with boosted regression trees.
The LHCb trigger was reoptimised, comparing the performance of several
tree-based algorithms to neural networks~\cite{Likhomanenko2015},
while their muon identification performance for the Run 3 of the LHC
will profit from improvements thanks to gradient
boosting~\cite{Anderlini:2020ucv}.

The latest result has just been published at the time of writing,
reporting the first evidence for $t\bar tt\bar t$ production in
ATLAS~\cite{TOPQ-2018-05}, shown in \fref{fig:BDTusage:tttt}.  The one
analysis using the most \bdts is probably the observation of the
diphoton decay of the Higgs boson by the CMS
experiment~\cite{CMS-HIG-13-001}. The diphoton vertex is selected with
a \bdt, while another one estimates, event-by-event, the probability
for the vertex assignment to be within 10~mm of the diphoton
interaction point. Photons are identified with a \bdt, and their
energy is corrected with a boosted regression tree that provides the
energy and its associated uncertainty. Finally several \bdts are used
to select the various signal regions and extract signal from these
different categories.

Beyond object identification and calibration, and final discriminant
in physics analyses, \bdts can also be used to reduce the number of
potential object combinations in order to find the correct match
between the observed objects in the detector and their probable source
of production. Such a `reconstruction BDT' was used to look for the
associated production of a Higgs boson and a pair of top quarks,
$t\bar{t} H(b\bar b)$~\cite{HIGG-2017-03}.

Lately there is a tendency towards deep neural networks and their many
flavours to replace \bdts
in the various stages of analysis~\cite{ATL-PHYS-PUB-2019-033,ATL-PHYS-PUB-2020-014}. \BDTs nevertheless remain a
favourite in high-energy physics, for their ease of use, high
performance out-of-the-box, limited required tuning of hyperparameters
and resilience against overtraining.

\subsubsection{Systematic uncertainties}
\label{sec:syst}
There is an \emph{a priori}, especially among physicists not very
familiar with machine learning techniques, to distrust their output
because they are not a measurable quantity with a physical meaning
like an invariant mass. They are indeed complex variables, but so are
for instance energy quantities for reconstructed particles in the
detector. Uncertainties on such `basic' variables are typically
evaluated by varying the value of a requirement, changing
the calibration of objects that go into the variable, etc. The \bdt
output (or of any such multivariate technique) is no different: its inputs can be varied according to their know uncertainties (for instance varying the jet energy scale will have a correlated impact on all discriminating variables that depend on jets) and
their effect propagated through the \bdt (the shifted inputs will lead to a different \bdt output), to see how much these changes impact the
analysis. This gives the size of the uncertainty on the multivariate
discriminant output.

That being said, the Peter Parker principle applies: `With great power
comes great responsibility'. \BDTs are very powerful, and will target
small areas of phase space where potentially not all known systematic
uncertainties are strictly valid. Then extra uncertainties may be
needed, not so much on the technique itself but rather due to the fact
that it extracts information from less well-known regions.

Usually \bdts are trained on the nominal Monte Carlo samples and are therefore
completely oblivious to the effect of systematic uncertainties. This
could lead to bad results once they are introduced, if the \bdts are
sensitive to them, and when applied on real data. One way to possibly mitigate this effect is with one form of data augmentation, training
the \bdts on a mixture of nominal and systematically shifted events, hence increasing the training statistics and allowing the \bdts to see other events than the nominal ones during training to learn their features. The
nominal performance should decrease, but with the hope that systematic
uncertainties will have less of an impact on the final
measurement. Experience with this approach is inconclusive.
If the physics model is not properly describing the real data, then the performance will also be affected. It can be partially addressed
with domain adaptation~\cite{BenDavid2009} (as described elsewhere in this book).

\section{Other averaging techniques}
\label{sec:Others}
As mentioned in \sref{sec:average} the key to improving a
single \dt\ performance and stability is averaging. Other techniques
than boosting exist, some of which are briefly described below. As with
boosting, statistical perturbations are introduced to randomise the training sample, hence increasing the
predictive power of the ensemble of trees.
\begin{description}
\item[Bagging] (Bootstrap AGGregatING) was proposed in
  Ref.~\cite{Bagging}. It consists in training trees on different
  bootstrap samples drawn randomly with replacement from the training
  sample. Events that are not picked for the bootstrap sample form an
  `out of bag' validation sample. The bagged output is the simple
  average of all such trees, with a reduced variance compared to
  individual trees.
\item[Random forests] is bagging with an extra level of
  randomisation~\cite{RandomForests}. Before splitting a node, only a
  random subset of discriminating variables is considered. The
  fraction can vary for each split for yet another level of
  randomisation.
\item[Trimming] is not exactly an averaging technique per se but can
  be used in conjunction with another technique, in particular
  boosting, to speed up the training process. After some boosting
  cycles, it is possible that very few events with very high weight
  are making up most of the total training sample weight. Events with
  very small weights may be ignored, hence introducing
  again some minor statistical perturbations and speeding up the
  training. $\varepsilon$-HingeBoost is such an algorithm (see
  \sref{sec:otherboost}).
\end{description}

\section{Software}
\label{sec:Soft}
Many implementations of decision trees exist on the market. Some of
them, all open source, are briefly presented below.

The most popular in high-energy physics is TMVA~\cite{TMVA},
integrated into ROOT. It includes single decision trees, boosted trees
with AdaBoost and gradient boost, bagging and random forests. Being
part of ROOT it is very straightforward to use within usual analysis
frameworks, both in C++ and Python. It includes tools for data
preparation and makes it simple to compare performance between many
algorithms, not only tree-based ones. Already mentioned
Refs.~\cite{ATLAS-CONF-2017-029,FTAG-2018-01,CMS-HIG-13-001,TOPQ-2018-05,HIGG-2017-03}
are but a few examples of TMVA usage in the field.

Another implementation has gained visibility in high-energy physics:
XGBoost~\cite{XGBoost}. It entered the field after receiving to
special HEP meets ML award during the Higgs boson machine learning
challenge (HiggsML) hosted by Kaggle~\cite{pmlr-v42-cowa14} (described
in Chapter 20). It features a high-performing, scalable
gradient boosting implementation, capable of using GPU and large
cluster parallelisation. Instead of the greedy algorithm described in
\sref{sec:algoDT}, the authors developed an approximate algorithm that
proposes candidate splitting points according to percentiles of the
input variables, and then maps the variables into buckets according to
these splits to find the best solution. Many analyses at the LHC are
now using it (see for instance Ref.~\cite{HIGG-2018-13}).

Other implementations have lower usage in high-energy physics so far
while being used in other fields. LightGBM (light gradient boosting
machine~\cite{LightGBM}), originally developed by Microsoft, is
competing with XGBoost in speed, scalability and performance. It
builds trees in a very different way from what was presented in this
chapter, with a histogram-based \dt learning
algorithm. Scikit-learn~\cite{scikit-learn} is a very popular machine
learning framework with several tree-related implementations and
utilities for data preparation. Finally CatBoost~\cite{CatBoost} is a
new gradient boosting implementation from Yandex used in commercial
services as well as in high-energy physics, for instance in
LHCb~\cite{Anderlini:2020ucv}.

\section{Conclusion}
\label{sec:Conclusion}
This chapter introduced what \dts\ are and how to construct them,
as a powerful multivariate extension of a cut-based analysis.
Advantages are numerous: their training is fast, they lead
to human-readable results (not black boxes) with possible
interpretation by a physicist, can deal easily with all sorts of
variables and with many of them, with in the end relatively few
parameters.

\DTs\ are, however, not perfect and suffer from the piecewise nature
of their output and a high sensitivity to the content of the training
sample. These shortcoming are for a large part addressed by averaging
the results of several trees, each built after introducing some
statistical perturbation in the training sample. Among the most
popular such techniques, boosting (and its AdaBoost and gradient boost
incarnations) was described in detail, providing ideas as to why it
seems to be performing so well while being very resilient against
overtraining.  Other averaging techniques were briefly presented.

Boosted decision trees have now become quite fashionable in high
energy physics. Following the steps of MiniBooNe for analysis and
particle identification and D0 for the first evidence and observation
of single top quark production, other experiments and analyses are now
using them routinely, in particular at the LHC.

\BDTs are still a very active field of development, with academic
groups and private companies testing their limits, providing new
software~\cite{CatBoost,XGBoost,LightGBM} and using them to target
recent issues like resistance to adversarial attacks (see
e.g.~\cite{NEURIPS2019_cd9508fd,NEURIPS2019_4206e389}).

\bibliographystyle{tepml}
\bibliography{Coadou-BDT}

\end{document}